\documentclass[ 
preprint,
superscriptaddress,
aip,
jcp,
 amsmath,amssymb,
floatfix,
]{revtex4-1}
\usepackage{dcolumn}
\usepackage{bm}
\usepackage{geometry}
\usepackage{graphicx} 

\usepackage{textcomp}
\usepackage{float}
\usepackage{booktabs} 
\usepackage{array} 
\usepackage{paralist} 
\usepackage{verbatim} 
\usepackage{subfig} 
\usepackage{rotating}

\usepackage{fancyhdr} 
\usepackage{threeparttablex}
\usepackage{amsmath, amsthm}
\usepackage{afterpage}
\usepackage[nottoc,notlof,notlot]{tocbibind} 
\usepackage[titles,subfigure]{tocloft} 

\graphicspath{{./epsfiles/}}

\usepackage{mathrsfs}
\DeclareMathAlphabet{\mathpzc}{OT1}{pzc}{m}{it}
\begin{document}

\title{Dimensional Scaling Treatment with Relativistic Corrections for Stable Multiply Charged Atomic Ions in High-Frequency Super-Intense Laser Fields}
\author{Ross~D.~Hoehn}
\affiliation{Departments of Chemistry and Physics, Purdue University,
West Lafayette, IN 47907 USA}
\author{Jiaxiang~ Wang}
\affiliation{State Key Laboratory of Precision Spectroscopy, Institute of Theoretical Physics, Department of Physics, East China Normal University, Shanghai China 200241}
\author{Sabre~Kais}
\thanks{Corresponding Author : kais@purdue.edu}
\affiliation{Departments of Chemistry and Physics, Purdue University,
West Lafayette, IN 47907 USA} 

\begin{abstract}
We present a theoretical framework which describes multiply-charged atomic ions, their stability within super-intense laser fields,also lay corrections to the systems due to relativistic effects. Dimensional scaling calculations with relativistic corrections for systems: H, H$^{-}$, H$^{2-}$, He, He$^{-}$, He$^{2-}$, He$^{3-}$ within super-intense laser fields were completed. Also completed were three-dimensional self consistent field calculations to verify the dimensionally scaled quantities. With the aforementioned methods the system's ability to stably bind 'additional' electrons through the development of multiple isolated regions of high potential energy leading to nodes of high electron density is shown. These nodes are spaced far enough from each other to minimized the electronic repulsion of the electrons, while still providing adequate enough attraction so as to bind the excess elections into orbitals.  We have found that even with relativistic considerations these species are stably bound within the field.  It was also found that performing the dimensional scaling calculations for systems within the confines of laser fields to be a much simpler and more cost-effective method than the supporting D=3 SCF method.  The dimensional scaling method is general and can be extended to include relativistic corrections to describe the stability of simple molecular systems in super-intense laser fields.

\end{abstract}

\maketitle
\section{Introduction}

The generation of stable, multiply-charged atomic ions via exposure to super-intense laser fields is a topic which challenges preconceived notions for ionic atoms and is, therefore, of fundamental importance in atomic and molecular physics\cite{0953-4075-35-18-201,Gavrila_Book,eberly}. Over the past decades, advancements in spectroscopic methods have yielded verification of mono-charged Calcium and Strontium atomic anions\cite{1987PhRvL..59.2267P}$^{,}$\cite{1995PhRvL..75..414B} and various gas-phase poly-charged molecular ions\cite{Scheller17111995}$^{,}$ \cite{SimonsII}$^{,}$ \cite{SimonsI} . However without the large charge volume which is provided by the heavy atoms -above- or small molecules it is unlikely that species would be able to bind more than one excess electron; this can be noted by the relative stability of O$^{-2}$ in the liquid-phase, yet it's instability within the gas-phase\cite{Scheller17111995}.Theoretical works have developed an absolute upper-limit to the number of electrons which may be bound to a atomic center\cite{1984PhRvA..29.3018L}: N$_{c}\geq 2$Z, with N$_{c}$ being the number of electrons and Z being the Coulomb charge of the nucleus. Within the context of Lieb's frameworks, Hydrogen would therefore be disallowed any excess electrons beyond that which yields the Hydride state, thusly H$^{2-}$ is unstable\cite{1987PhRvA..35.3949C}$^{,}$\cite{Bates19901}. Supporting theoretical works have come later\cite{0305-4470-32-39-312}$^{,}$\cite{Serra2003205} -some including implementation of finite-sized scaling\cite{Serra2003205}- and have conclusively determined at gas-phase, dianionic atoms are unstable. 

It has been shown that stable, multiply-charged atomic ions may be developed within extremely strong laser fields on the order of 10$^{16}$ W/cm$^{2}$ and above\cite{2007JChPh.127i4301W}$^{,}$\cite{2006JChPh.124t1108W}.
Within the field, the electron density - still being bound to the nucleus - has been found to be nodal in nature as the Coulomb potential splinters under the influence of the field into distinct, localized regions whose positions are governed by the field parameters of the laser. This phenomenon is most easily - and best- discussed within the context of the Krammers Henneberger (KH) reference frame, electron centric frame, where the electron is treated as the stationary body and the nucleus traverses the path of the applied field; in this context the local nodes of electron density are located at the turning points on the path of the nucleus. These are the location at which the angular velocity of the nucleus decreases and thus spends more time in a local area - thus generating a greater pull in that area. Within these nodal regions, the bound electrons maintain a great enough
distance from one and another to minimize their Coulomb repulsion while also giving each a “center” with which to bind. In this field, the electrons -which
intuition tells us would be completely ionized- are capable of stably binding into multiply-charged atomic ions. The field strength allows one to manipulate the location and pull of the nodal centers , thus generating a method of control over the potential and therefore establishing the ability to push the electrons into and past their most stable state by means of
manipulating laser parameters, frequency and intensity.

The contained theoretical works are concerned with High-Frequency Floqeut Theory (HFFT) which allows for a  time-independent  treatment of the coupling of the static
 Coulomb potential with a time-varying electromagnetic field.  This is possible by exploiting highly oscillatory fields in which the electrons would be prohibited from
 coupling with the periodic nature of the field due to extremely short periods (large frequencies) of oscillation, thus the system's electrons would feel a period average 
of the applied potential coupled with the static Coulomb potential, again this is best discussed within the KH frame.  This time-average allows  generation of the aforementioned nodal structure and therefore
 permits the stability of the subsequent states and allows the system to forgo autoionization.  The above discussed methodology was introduced to atomic systems by
 Pont et al\cite{PhysRevLett.61.939}, van Duijin et al\cite{vanDuijin}, and was used again by Wei et al\cite{2006JChPh.124t1108W}$^{,}$\cite{PhysRevA.76.013407}$^{,}$\cite{2008JChPh.129u4110W}$^{,}$\cite{2007JChPh.127i4301W} to describe non-relativistic, multiply-charged atomic ions.  Herein we shall propose a framework utilizing HFFT as a
 backbone for applying relativistic corrections to atomic ions in a time-independent manner.

\section{Relativistic Corrections}
\label{Rel Cors}

\subsection{Non-Relativistic Methodology}

Consideration within the non-relativisitic cases lies no longer with both the mass and the magnetic coupling, but with the time dependent electric field coupling with the system's Coulomb potential; this work was performed by Wei et al\cite{2006JChPh.124t1108W}$^{,}$\cite{PhysRevA.76.013407}$^{,}$\cite{2008JChPh.129u4110W}$^{,}$\cite{2007JChPh.127i4301W} and produced stably bound multiply-charged ions for small atomic centers utilizing the field parameter ($\alpha_{0}$), discussed later, and finding detachment energies on orders of 0.1eV to  1.0 eV.  The enclosed works, here, expound upon this by adding the necessary relativistic corrections to the previous framework.  A free electron within an oscillating electric field shall undergo oscillatory motions which are governed by a coupling to the field; the electron is said to be 'quivering' with a motion defined by a trajectory, $\vec{\alpha}_{0} (t)$, and a  quiver amplitude, $\alpha_{0}$.  A bound electron within the same situation shall feel a new potential which is a stacking of the applied field and the Coulomb potential of the central charge; the total potential for the system is said to be a Coulomb potential dressed by the laser, denoted as a dressed potential, $V_{dres}$.  

Under the auspices of HFFT, introduced above and here\cite{PhysRevLett.61.939}, by applying a highly oscillatory laser field with an extremely short period the electrons will lack the ability to oscillate synchronously with the the applied field. In this manner, the potential felt by the electrons is a period average of the oscillatory field, this new potential is a dressed potential under the HFFT approximation, $V_{dres}^{HFFT}$. In all cases addressed within this paper the laser-coordinates are: laser fired in y-direction, electronic component linearly polarized in the z-direction, and the magnetic component in the x-direction

The situation of the dressed potential,$V_{dres}^{HFFT}$, is a time independent problem as the field has been period averaged, due to this the full Hamiltonian can be treated within the Time Independent Schrodinger Equation (TISE):
\begin{equation}
\epsilon_{i}\Psi_{i}=\hat{H} \Psi_{i}=\frac{-\hbar^{2}}{2 m_{e}}\nabla^{2}\Psi_{i}+V^{HFFT}_{dres}\Psi_{i}.
\end{equation}
Accurate solutions to the equation are difficult for systems of more than one electron due to many-body interactions, but approximate solutions can be obtained in a self-consistent method (SCF) through Hartree-Fock (RHF/UHF), Density Functional Theory (DFT), or post-Hartree-Fock Methods.

\subsection{The Relativistic Mass Gauge}

Mass -as a fundamental- is conceptualized in two different manners within physics, these being the rest mass and the relativistic mass. Rest mass, or invariant mass, is for a specific body a constant, measurable
quantity denoted m$_{0}$. In opposition to the rest mass is the variant quantity: relativistic mass, m$_{r}$. The relativistic mass depends upon the velocity of the observer.  The variant nature of $m_{r}$ is a correction to the rest mass which accounts for a non-zero kinetic energy for the measured system.
This means that the relativistic mass increases in magnitude as the velocity of the system increases, and shall reach infinite mass as the system reaches the speed of
light.

This portrayal of the mass shall be implemented within the Time Independent Schrodinger Equation (TISE) for the enclosed work,  we shall now need to express alterations
 to the rest mass in terms of the system's laser parameters. 

We are now called to introduce the concept of ponderomotive energy, U$_{p}$; this being the cycle average
kinetic energy of a quivering electron, i.e. electron undergoing oscillatory motion due to an external field and also qualifying under the dipole approximation. This is quantity discussed in context of such systems by Joachain, D$\ddot{o}$rr and Klystra\cite{Joachain2000225}.
\begin{equation}
U_{P} = \frac{e^{2}\mathscr{E}^{2}_{0}}{4m_{e}\omega^{2}}
\end{equation}
In the above, $\mathscr{E}_{0}$ is the peak strength of the electric field, $\omega$ is the angular frequency of the applied field and both e and m$_{e}$ retain their conventional meanings: of magnitude of electron charge and electron mass, respectively. This quantity aides in the evolution of the rest mass to the relativistic mass as:
\begin{subequations}
\begin{eqnarray}
m_{r}= m^{dressed} &=& m_{e}(1+2\frac{U_{P}}{m_{e}c^{2}})^{\frac{1}{2}} \\
&=& m_{e}(1+2q)^{\frac{1}{2}}.\label{thinger}
\end{eqnarray}
\end{subequations}
As can be seen, the quantity \emph{q} begins to shift the mass and becomes the dominant factor within the expression as it approaches unity\cite{2003PhyS...68C..72J}. The form of Eq. (\ref{thinger}) was found by Brown and Kibble\cite{PhysRev.133.A705} and later verified by Eberly and Sleeper\cite{PhysRev.176.1570} via the Hamilton-Jacobi equation.

Below we shall discuss the above mass concept as it applies to the TISE for one electron (which can then be generalized to N electrons), these lines shall be discussed stepwise. 
\begin{eqnarray}
Eq. (1) &\rightarrow&\frac{-\hbar^{2}}{2 m_{e}(1+2q)^{\frac{1}{2}}}\nabla^{2}\Psi_{i}+V\Psi_{i}\label{lineone}\\
&\rightarrow&\frac{-\hbar^{2}}{2 m_{e}(1+2\frac{e^{2}\alpha^{2}\omega^{2}}{4m_{e}^{2}c^{2}})^{\frac{1}{2}}}\nabla^{2}\Psi_{i}+V\Psi_{i}\\
&\rightarrow&\!\!\frac{-\hbar^{2}}{2(1\!\!+\!\!2.66\!\!\times\!\! 10^{-5}\alpha^{2}\omega^{2})^{\frac{1}{2}}}\nabla^{2}\Psi_{i}\!\!+\!\!V\Psi_{i}\label{linethree}
\end{eqnarray}
 The first line, Eq. (\ref{lineone}), shows the form of the TISE as it appears accounting for the mass gauge, which is tuned by the quantity \emph{q}.  Secondly, we have introduced and employed the field coefficient, $\alpha=\frac{\mathscr{E}_{0}}{\omega^{2}}$, as a means of defining \emph{q} in terms of known laser parameters.  Lastly, we express all quantities in $\hbar$=m$_{e}$=1
 units (atomic units), this allows us to maintain the relativistic alterations as a unitless multiplier,  Eq. (\ref{linethree}).  As the multiplier which transforms invariant mass to relativistic mass is a unitless quantity, the
 resultant energies from the final line, Eq. (\ref{linethree}), shall be in Hartree E$_{H}$, as they would if one ignored the mass gauge entirely.  In all cases considered within this paper the potential function, V, shall be dressed under HFFT , making V=$V^{HFFT}_{dres}$ 

\subsection{Trajectory Corrections}

High-Frequency Floquet Theory (HFFT) was first introduced to similar systems by Pont et al\cite{PhysRevLett.61.939}, relies on the frequency of the externally applied electromagnetic field to be so quickly oscillating so as even the electrons are incapable of coupling their motions to the field.  In this manner the D-dimensional dressed Coulomb potential -which is in essence a time-dependent problem– is simplified to a time-independent problem:
\begin{eqnarray}
\!\!\!\!\!V_{dres}^{HFFT}\!\!=\!\frac{Z}{2 \pi} \int\limits_{0}^{2\pi}\left(\frac{\,\, d(\omega t)}{ \!\sqrt{\sum\limits_{i}^{D}{(x_{i} + \alpha_{i})^{2}}}\!}\right).
\end{eqnarray}
To apply the above period average to a system one must develop an interest in the trajectory, $\vec{\alpha}(t)$, of the laser's path as it's components are required in the above averaging as $\alpha_{i}$ along with the i$^{th}$ Cartesian component, $x_{i}$.  Earlier works\cite{2007JChPh.127i4301W}$^{,}$\cite{2006JChPh.124t1108W}$^{,}$\cite{PhysRevA.76.013407}$^{,}$\cite{2008JChPh.129u4110W} have concentrated upon non-relativistic systems, and thus the laser trajectory is equivalent to the path taken by a free electron undergoing influence by a time-dependent external electric field (or laser field where no magneto-coupling is considered);  the up and down oscillatory motion of the electric field governs the trajectory of the electron, seen in Eq.(\ref{eq:non-realtraj}), where the polarization is as discussed in $\S$A.
\begin{subequations}\label{eq:non-realtraj}\begin{eqnarray}
\!\!\!\!\vec{\alpha}(t)&=&\langle \alpha_{x}, \alpha_{y}, \alpha_{z} \rangle\\
& =& \langle 0, 0, \alpha_{0}cos(\omega t) \rangle .
\end{eqnarray}
\end{subequations}
Now concerning ourselves with the electronic-magnetic coupling within relativistic regimes; this shall be described analogously to the non-relativistic case above by the path taken by a free electron in an electromagnetic field now with the electronic-magneto coupling accounted.  Within a plane-wave laser field, the electron classical trajectory can be obtained analytically. For a linearly polarized laser field, the non-zero field components can be written as:
\begin{eqnarray}
E_z=E_0 cos\eta,\\
cB_x=E_0 cos\eta,
\end{eqnarray}
where $\eta=\omega t-ky$ is the phase of the field. By using Newton-Lorentz equation,
\begin{equation}
\frac{d\vec P}{dt}=-e(\vec E+\vec v\times\vec B),
\end{equation}
we have,
\begin{equation}
\begin{split}
\frac{d\vec P}{dt}&=\{\frac{dP_x}{dt}, \frac{dP_y}{dt},  \frac{dP_z}{dt}\}\\
&=\{-eE_0\beta_xcos\eta, 0,-eE_0(1-\beta_z)cos\eta\} \label{eqp}
\end{split}
\end{equation}
\begin{equation}
\!\!\!\! {m_e}{c^2}\frac{{d\gamma }}{{dt}} =  - e{E_0}\cos \eta  \cdot {v_x},
\label{eqenergy}
\end{equation}
where $\gamma=\sqrt{1+{\vec P}^2/(m_e^2c^2)}$ is the Lorentz factor.
For an electron initially at rest, from Eqs.(\ref{eqp}-\ref{eqenergy}), we obtain:
\begin{equation}
\begin{split}
\vec P& = \{P_x, P_y, P_z\}\\
& = \{\frac{e^2E^2_0}{2m_ec\omega^2}sin^2\eta, 0, -\frac{eE_0}{\omega}sin\eta\}
\end{split}
\end{equation}
From the above an electron's trajectory can be acquired,
\begin{eqnarray}
\vec \alpha=\{-\frac{c}{8\omega}Q_0^2sin(2\eta), 0 , \frac{c}{\omega}Q_0cos\eta\}.
\end{eqnarray}
Where $Q_0=eE_0/(m_e\omega c)$ and electron drift motion has been neglected. Within dipole approximation,
\begin{equation}
\vec \alpha = \{0,-\frac{c}{8\omega}Q_0^2sin(2\omega t), \frac{c}{\omega}Q_0cos( \omega t)\},
\end{equation}
which means:
\begin{equation}
\begin{split}
\frac{d^{2} \vec \alpha}{dt^{2}} &= \{\ddot \alpha_x, \ddot \alpha_y, \ddot \alpha_z \}\\
&=\{\frac{1}{2}c\omega Q_0^2sin(2\omega t),  0, -c\omega Q_0 cos(\omega t) \}.\label{eq:ddotalpha}
\end{split}
\end{equation}
Eq. (\ref{eq:ddotalpha}) tells us that we can approximately take the above relativistic trajectory as an equivalent one for the electron moving in the following effective electric fields,
\begin{eqnarray}
\!\!\!\!\!\!\vec E\!\! =\!\! \{\!\frac{\omega m_ec}{2e} Q_0^2sin(\!2\omega t\!)\!,  0,\!\! -\frac{\omega m_e c}{e}Q_0 cos(\!\omega t\!)\! \},
\end{eqnarray}
which will be used in HK theory. In atomic units, the trajectory can be written as,
\begin{eqnarray}
\vec \alpha  = \{-\alpha_0^2\alpha_f sin(2\omega t),  0, \alpha_0 cos( \omega t)\}
\end{eqnarray}
By comparing with the non-relativistic trajectory, we have an extra oscillation motion along the laser propagation direction, which comes from the
magnetic coupling, also intorduced is the fine structure constant which mediates the magnetic-electronic coupling term, $\alpha_{f}$. This results in the famous figure-8 motion. To illustrate the effects of this correction in HK effective potential, we will first take a 1-D box potential as an example in the following.

\subsection{1-D Particle in a Box}
For simplicity, we take the 1-D box potential as follows,
\begin{equation}
V(z)=\left\{
\begin{array}{l}
\pi,\rm{   }\left\vert z\right\vert \le 1,\\
0, \rm{ }\rm{elsewhere}.
\end{array}
\right.
\end{equation}
Then the HK effective potential can be acquired analytically by the following integration.
\begin{equation}
V_{eff}(z)=\frac{1}{2\pi}\int_0^{2\pi}V(z+\alpha_0cos(\Omega))d\Omega.
\end{equation}
When $0\leq\alpha_0 \leq\frac{1}{2}$,\\
\begin{equation}
V_{eff}(z)=\left\{
\begin{array}{l}
arcos\left[-\frac{z+1}{\alpha_0}\right], -\alpha_0\leq z\leq-1+\alpha_0,\\
arcos\left[\frac{z-1}{\alpha_0}\right], 1-\alpha_0\leq z\leq 1+\alpha_0,\\
\pi,\:\:-1-\alpha_0\leq z \leq 1-\alpha_0,\\
0, \rm{elsewhere}.
\end{array}
\right.
\end{equation}
When $\alpha_0\geq\frac{1
}{2}$,
\begin{equation}
V_{eff}(z)=\left\{
\begin{array}{l}
arcos\left[-\frac{z+1}{\alpha_0}\right], -\alpha_0\leq z\leq 1-\alpha_0,\\
arcos\left[\frac{z-1}{\alpha_0}\right], 1-\alpha_0\leq z\leq -1+\alpha_0,\\
arcos\left[\frac{z-1}{\alpha_0}\right], -1+\alpha_0\leq z\leq 1+\alpha_0,\\
0, \rm{elsewhere}.
\end{array}
\right.
\end{equation}

Fig.\ref{nonrelativistic} shows how the box potential is modified by the external laser field. It is clear to see that as the laser intensity increases above certain value, the original potential
will evolve into a double well. Moreover, the two wells will become more separate and more shallow if $\alpha_0$ is further increased. This indicates two important features for the ground state in this potential. Firstly, the electrons will become less bound or the potential might have higher ground state energy. Secondly, if we have two electrons in this effective potential, they will have more space in which to avoid each other, which means the electron repulsion energy will tend to be smaller for un-paired electrons. Hence the final ground state energy for multi-electron systems should depend upon the competition of these two factors.

Once we introduce the relativistic corrections to the  electron trajectory, the effective potential along $z$-axis will becomes:
\begin{equation}
\label{eq:zcorrect}
V_{eff}\!(z)\!\!=\!\!\frac{1}{2\pi}\!\!\int_0^{2\pi}\!\!\!\!\!\!\!\!d\Omega\!\!\left(\!\!\sqrt{\!(z\!+\!\alpha_0cos\Omega)^2\!+\!\alpha_0^2\alpha^2sin^22\Omega}\!\right),
\end{equation}
in which the integral will be calculated numerically.
From Fig.(\ref{relativistic}),it is interesting to note that, when $\alpha_0$ is large enough, for example $\alpha_0=10$ as in Fig.\ref{relativistic}(b), the effective potential will have three local minima.  Qualitatively, this should result from the relativistic figure-8 motions, which helps the electron maintain a position nearer the orbit center for a greater period of time. Another characteristic is that the three separate minima become much shallower for higher laser intensity. These drastic changes over the effective potential will be expected to have influences over the many-electron states bound by the potential. It seems that the relativistic effect provides us another way to engineer the potential. Based upon the observations over Eq.(\ref{eq:zcorrect}), we can even think about using lasers with different colors superposed together to have more freedoms modifying the effective potentials. Fig. \ref{multicolor} shows us one example with two alternating electric fields as follows,
\begin{equation}
\vec \alpha(t)=\alpha_0cos(\omega t)\hat e_x+\alpha_1 sin(4\omega t)\hat e_z.
\end{equation}

\subsection{Potential under Consideration}

Beginning with the spherically symmetric Coulomb potential, the applied external fields shall morphologically alter this potential to conform to the trajectory discussed in $\S$ B.  In doing this there are three main regimes in which the potential may exist: firstly, the spherical; secondly, the pseudo-linear; and finally, the parametric.  The potential only exists within the spherical regime when there is no applied external field, as the field begins to evolve the coupling of the external field and the Coulomb potential becomes apparent with the electric field component dominating; this creates a regime where the system maintains an almost linear behavior as if there were no relativistic correction to the trajectory.  Distortion of the pure linear nature exists but is a small effect compared to the primary electric effect.

 As the intensity of applied field increases the magnitude of the magnetic field begins to compensate for the dampening effect of the fine structure constant, $\alpha_{f}$; as this takes place, the magnetic contributions to the field coupling begin to dominate the system, whose character now exhibits the hourglass figure of the parametric regime.  Figure \ref{fig:contours} shows a series of contour plots of the potential energy plotted in the x-z directions for a series of field intensities, $\alpha_{0}$; the behavior of the system and it's development through the previously discussed regimes should be obvious.
A three-dimensional plot of the potential energy surface for \begin{math}\alpha_{0}= 100 \end{math} is also enclosed as Fig. \ref{fig:surfaces}, this plot only considers displacements in the spatial x-z directions for reasons introduced in $\S$ A, where the vertical axis describes the magnitude of the potential energy at this x-z location.

This potential is ideal for attempting to develop mutliply-charged ionic system from small nuclei, as it maintains a potential well at the center of the systems along with four other locations along the parametric trajectory; this allows bound electrons to attach to the individual potential wells while maintaining a large enough distance from each other to minimize electron-electron repulsions.

\section{Methodology for D=3 Calculations}
\label{Sec: 3-D}
Three dimensional calculations which describe our systems were executed as a mean of verifying the simpler Dimensional Scaling approach discussed later. The methodology consisted of unrestricted Hartree-Fock (UHF) utilizing the Pople-Nesbot equations -which allow for the accommodation of basis sets- to complete calculations for a series of total electron counts, N, per single value of the field intensity parameter, $\alpha_{0}$. The calculations were dependent both upon appropriate choice of basis set and upon the locations selected as the centers for these basis sets; for the determinations of the locations of probable electron density we deferred both to 'exact' locations of the electrons from the limit D$\rightarrow \infty$ calculations and to contour plots of the potentials for a given field intensity value, see Figure \ref{fig:contours} for an example.
Upon discerning from the above information the locations of the psuedo-centers within the system-space, a basis set was selected which could describe these nodes. There exist, at minimum, 5 distinct nodes of electron density within the system, these being at the center –coexisting with the origin of our coordinates systems- and at four psuedo-centers residing upon the parametric curve described by the relativistic trajectory used within the HFFT Potential.
\begin{eqnarray}
\left\{
\begin{array}{ll}
x\\
y\\
z
\end{array}
\right\}
=
\left\{
\begin{array}{ll}
-( \frac{\alpha_{0}^2}{\alpha_{f}} ) sin (2 t ) \\
0\\
\alpha_{0} cos ( t)
\end{array}
\right\}
\end{eqnarray}
The orbital centers were selected to satisfy the above curve and to coincide with the "hottest" locations displayed within the contour plots of potential energy. A basis set was selected which allowed for significant description of both polarized and diffuse phenomenon residing on small centers. The centers of potential electron density which do not reside at origin were described by basis sets optimized to describe the atom of the nucleus at the system's origin, i.e. all five centers on hydrogen use a hydrogen-fit basis set.
Overlap and kinetic integrals were performed with variations on the integrals described by McMurchie and Davidson in their seminal paper here\cite{McMurchie1978218}. Numerical integration methods were appropriate for the more challenging, non-analytic potential integrals. A global adaptive method was applied to the system which performs admirably with Gaussian functions placed at the coordinate origin; but as the method relies on sampling points within the equation-space to find non-zero areas of the function, Gaussians placed distances away from the origin were sometimes so small relative to the distance from origin as to be invisible. For this reason two-centered intregral includes the cost of shifting the coordinate-origin to the center of the product Gaussian as defined by\cite{Szabo_Ostlund_1996}: 
\begin{equation}\label{gaussprod}
\begin{split}
\phi^{GF}_{1s}& (\alpha_{A}, r\!\!-\!\!R_{A})\!\! \times\!\! \phi^{GF}_{1s} (\alpha_{B}, r\!\!-\!\!R_{B}) \\
&= exp \{ -(\alpha_{A}(r\!\!-\!\!R_{A}) + \alpha_{B}(r\!\!-\!\!R_{B})) \} \\
& = exp\{ -\alpha_{a} \alpha_{B}(\alpha_{A}\!\!+\!\!\alpha_{B})^{-1}|R_{A}\!\!-\!\!R_{B} |^{2}\}\times\\
&\quad \quad\quad \phi^{GF}_{1s}(\alpha_{A}\!\!+\!\!\alpha_{B}, r-R_{P}).
\end{split}
\end{equation}
In this way the chosen method of numerical integration was capable of adequately describing the three-dimensional potential energy integrals. Single electron cases were verified prior to enacting the self-consistentt field calculations, as the energies of the single electron system may be revealed as the eigenvalues of the canonically orthogonalized\, H$^{core}$ matrix alone, H$^{core}_{\mu, \nu} = $T$_{\mu, \nu}$+V$_{\mu, \nu}$. The four-centered integrals needed for the self-consistency were generated by exploiting the axillary function defined by Boyes\cite{1950RSPSA.200..542B}. The calculations for the mutli-electron energies were performed self-consistently with a convergence set to six decimals of accuracy, as chemical accuracy is define as 1.6mE$_{H}$, this set limit should suffice. 

 A plot of the square of the linear combination of atomic orbitals which comprised the set describing the appropriate eigenvalue yields semblance to the wave function of the system, whose probability density$ (|\Psi|^{2}) $ is shown in Figure \ref{fig:lastpix} for H$^{-}$ and for He$^{-}$, a two and a three electron case. It should be noted, as the D=3, UHF calculations were employed to verify the results of the Dimensional Scaling approach, that the locations of electron density shown in the aforementioned probability density plots speak to the validity of the Dimensional Scaling approach as the locations of the delta functions, see discussion provided in Results Section, which describe the seemingly stationary locations of the electrons in the limit of D $\rightarrow \infty$ calculations are very near those locations of highest probability density for finding elections given by the D=3 calculations.
Three-dimensional verification was performed for both the $H^{-}$ and $He^{-}$ species, as the regions of high potential energy become very delocalized for the remaining high-field species.  Aside from the high level of delocalization prohibiting the species from being described sufficiently with an appropriate number of Gaussians, the potential also spans a region of space on the order of 100 Bohr radii, yet optimized Gaussian basis sets for atomic centers span an order of 1-10 Bohr radii; this prohibits overlap for these species forcing the use of more and more Gaussians.  This acts prohibitively as the matricies required for UHF calculations are n$\times$n, where n scales as the number of centers by the number of basis contractions;  this obviously limits the achievable intensities applied to systems which can be calculated in this manner, especially on stand alone machines.

\section{Dimensional Scaling: Calculations and Considerations}
\label{Sec:D-Scaling}

\subsection{Methodology}

Many Body interactions are something which has troubled computational methodologies within quantum mechanics since inception; throughout the years the physical and chemical communities have made great advances in the field of electronic structure theory to help account for these electron-electron interaction through variational practices such as the Hartree Fock Method or Density Functional Theory. The alternative method to the aforementioned is a Dimensional Scaling treatment pioneered by Herschbach \cite{DudsPap}, discussed in\cite{Herschbach,Bohrs,sabre,Dunn,Avery}, and is briefly introduced here for the central force problem\cite{Herschbach}. Given the TISE for the simple central force problem in D-dimensions:
\begin{equation}
\label{eq:CFSE}
[ -\frac{1}{2}\nabla^{2}_{D}+V(r) ]\Psi_{D}=\epsilon_{D}\Psi_{D}.
\end{equation}
If we were to employ polar coordinates to the above we would require:
\begin{subequations}
\begin{eqnarray}
r\equiv [\sum\limits^{D}_{k=1}x^{2}_{k}]^{\frac{1}{2}}\label{eq:rdef} \\
\nabla_{D}^{2}=\frac{1}{r^{D-1}}\frac{\partial}{\partial r}(r^{D-1}\frac{\partial}{\partial r})-\frac{L^{2}_{D-1}}{r^{2}}.\label{eq:Laplacdef}
\end{eqnarray}
\end{subequations}
Where Eq.(\ref{eq:rdef}) gives the definition of the radial coordinate in a generic D-scaled space, and Eq.(\ref{eq:Laplacdef}) is the polar Laplacian in this D-scaled space, L$^{2}_{D}$ is the term which retains all angular dependencies. These angular and radial terms shall be dealt with in an divide and conquer treatment reminiscent to the radial and angular terms of the Rigid Rotor/Harmonic Oscillator approximations for the simple diatomic. We first write the wave function in D-dimensions to be the product: $\Psi_{D}=r^{\mathpzc{l}}\mathcal{Y}(\Omega_{D-1})$, where all radial dependencies are in the r$^{l}$ term and the D-1 remaining angular dimensions are described through $\mathcal{Y}(\Omega_{D-1})$. Now solving the angular terms for the form Eq.(\ref{eq:RigRot}), and the recognizing that the V(r) term in Eq.(\ref{eq:CFSE}) can be set to equal magnitude as the $\epsilon_{D}$ term, thus making Eq.(\ref{eq:CFSE}) reduce to the Laplace equation shown in Eq.(\ref{eq:LaplaceEq}).
\begin{equation}
L^{2}_{D-1}\mathcal{Y}(\Omega_{D-1})=C\mathcal{Y}(\Omega_{D-1})\label{eq:RigRot}\\
\end{equation}
\begin{subequations}\label{eq:LaplaceEq}
\begin{eqnarray}
\nabla^{2}_{D}r^{\mathpzc{l}}\mathcal{Y}(\Omega_{D-1})=0\\
\{\mathpzc{l}(\mathpzc{l}+D-2)-C\} r^{\mathpzc{l}-2}\mathcal{Y}(\Omega_{D-1})
\end{eqnarray}
\end{subequations}
This means: C=$\mathpzc{l}(\mathpzc{l}+D-2)$; and the Hamiltonian Operator in Eq.(\ref{eq:CFSE}) is now of the form:
\begin{equation}
\hat{H}_{D}=-\frac{1}{2}K_{D-1}(r)+\frac{\mathpzc{l}(\mathpzc{l}+d-2)}{2r^{2}}+V(r)\label{eq:NHO}.
\end{equation}
In the above, K$_{D-1}$(r) is the single non-angular term from the polar Laplacian in Eq.(\ref{eq:Laplacdef}).  We may now pass the system through a unit Jacobian, making: $J_{D}|\Psi_{D}|^2=\Phi_{D}$, J$_{D}$ is the radial part of the unit Jacobian and is of the form: r$^{(D-1)}$.  This would mean, $\Psi_{D}=r^{-\frac{1}{2}(D-1)}\Phi_{D}$. All leading to the form of K$_{D-1}$ seen here:
\begin{equation}
\label{eq:KD}
K_{D-1}(r)\!\!=\!\!r^{-\frac{1}{2}(D-1)}\{\frac{\partial^{2} \Phi_{D}}{\partial r^{2}}\!\!-\!\!\frac{D\!\!-\!\!1}{2}\frac{D\!\!-\!\!3}{2}\frac{\Phi_{D}}{r^{2}}\}
\end{equation}

Reassembling all the above, and placing them appropriately back into Eq.(\ref{eq:CFSE}), one shall -after menial simplification- get:
\begin{equation}
\{-\frac{1}{2}\frac{\partial^{2}}{\partial r^{2}}+\frac{\Lambda(\Lambda + 1)}{2r^{2}}+V(r)\}\Phi_{D}=\epsilon_{D}\Phi_{D}.\label{eq:RDSE}
\end{equation}
Eq.(\ref{eq:RDSE}) is the radial, D-scaled form of Eq. (\ref{eq:CFSE}), where the only dimensional dependencies lay within the $\Lambda$ terms as: $\Lambda = \mathpzc{l} + \frac{1}{2}(D-3)$.  The above leads to the minimization problem defined by the Hamiltonian discussed in $\S$B.

\subsection{Planar Infinite-D Hamiltonian}

Prior works published\cite{2008JChPh.129u4110W}$^{,}$\cite{2007JChPh.127i4301W} have also described systems both by infinite dimensional limit and then verified with three dimensional self consistent methods.  The dimensional scaled Hamiltonian presented in previous works was diatomic in nature and of the form\cite{avery_book}:
\begin{equation}
\label{eq:diatomic}
\begin{split}
\mathcal{H}_{DA}=&\frac{1}{2}\sum\limits^{N}_{i=1}\frac{1}{\rho_{i}^{2}}+\sum\limits_{i=1}^{N}V(\rho_{i}, z_{i})\\
&\quad+\sum\limits_{i=1}^{N}\sum\limits_{j=i+1}^{N-1}\!\!\frac{1}{\sqrt{(z_{i}\!\!-\!\!z_{j})^{2}\!\!+\!\rho_{i}^{2}\!\!+\!\!\rho_{j}^{2}}}
\end{split}
\end{equation}
and relied on previous works in which D-scaled Hamiltonians for diatomic systems were constucted, these diatomic Hamiltonians are also of the form above, and denoted $\mathcal{H}_{DA}$.

Hamiltonians of this form are applicable to the previous works as those non-relativistic systems,this is due to the consideration in absence of the second degree of symmetry breaking in the linear potential systems.

The diatomic-based Hamiltonian performed well on linear systems, but when attempting to use the above described Hamiltonian on a relativistically corrected trajectory, it was found that the equation behaved erratically with a smoothly evolving intensity as seen in Figure \ref{fig:Global}, and was in disagreement with the early 3-D calculations.  Although this Hamiltonian does not work well, overall it was found to be in good agreement while the potential was in the spherical and pseudo-linear regimes, failing only as the system approached and entered the parametric regime.

As dimensionally scaled Hamiltonians are not unique in nature, they are not singular in form.  We relied on arguments based on the breaking of spherical and then cylindrical symmetries to generate the followed assumed Hamiltonian for systems with broken radial and cylindrical symmetries, yet maintaining three orthogonal planes of symmetry.

As the dimensional-scaled, single electron central force problem yields as it's Hamiltonian $\mathcal{H}_{CF}$,
\begin{equation}
\label{eq:central}
\mathcal{H}_{CF}=\frac{1}{2 \:  r^{2}}+V(r).
\end{equation}
One can see from above the spherical nature of all terms within the Hamiltonian, as the potential is radial.  This equation predicts the ground state Hydrogen energy to be at $-\frac{1}{2}E_{H}$, exactly where it should be, and predicts the inter-atomic distance between the electron and the proton to be 1 in units of R$_{Bohr}$.  If one were to alter the above potential to conform with either the relativistic or non-relativistic cases discussed in this paper, the energies obtained would possess no physical significance and would overall behave similarly to the diatomic equation with the relativistic trajectory.  The addition of multiple electrons to the spherical symmetric problem yields :
\begin{equation}
\label{eq:multieleCF}
\begin{split}
&\mathcal{H}_{CF}=\frac{1}{2}\sum\limits^{N}_{i=1}\frac{1}{r_{i}^{2}}+\sum\limits^{N}_{i=1}V(r_{i})\\
&\quad+\sum\limits^{N}_{i=1}\sum\limits^{N-1}_{j=i+1}\frac{1}{\sqrt{r_{i}+r_{j}}}.
\end{split}
\end{equation}
Now we examine the diatomic case shown in Eq. (\ref{eq:diatomic}).  This equation has been designed to allow for potential which are of a cylindrical nature, as it must for a diatomic system.  Reevaluating the equation in Cartesian coordinates yields:
\begin{equation}
\label{eq:diatomicCC}
\begin{split}
\mathcal{H}_{DA}=&\frac{1}{2}\sum\limits^{N}_{i=1}\frac{1}{x_{i}^{2}+y^{2}_{i}}+\sum\limits_{i=1}^{N}V(x_{i}, y_{i}, z_{i})\\
&\quad+\sum\limits_{i=1}^{N}\sum\limits_{j=i+1}^{N-1}\!\!\frac{1}{\sqrt{(z_{i}\!\!-\!\!z_{j})^{2}\!\!+\!\!x_{i}^{2}\!\!+\!\!x^{2}_{j}\!\!+\!\!y^{2}_{i}\!\!+\!\!y^{2}_{j}}}.
\end{split}
\end{equation}
Reevaluation of the spherically symmetrical case in Cartesian coordinates gives:
\begin{equation}
\label{eq:multieleCC}
\begin{split}
&\mathcal{H}_{CF}=\frac{1}{2}\sum\limits^{N}_{i=1}\frac{1}{x_{i}^{2}+y_{i}^{2}+z_{i}^{2}}+\sum\limits^{N}_{i=1}V(x_{i}, y_{i}, z_{i})\\
&\quad+\sum\limits^{N}_{i=1}\sum\limits^{N-1}_{j=i+1}\frac{1}{\sqrt{x_{i}^{2}+x_{j}^{2}+y_{i}^{2}+y_{j}^{2}+z_{i}^{2}+z_{j}^{2}}}.
\end{split}
\end{equation}

As can been seen from the above, the orthogonal coordinate which is being cleaved from spherical symmetry has been removed from the kinetic evaluation and have been treated as a difference in the electron-electron term.  Continuing to use Cartesian coordinates, as it is this coordinate system which makes the relations apparent, we can move to an equation where the x-coordinate is now allowed to deviate from radial symmetry.  This again would remove the symmetry breaking coordinate from the kinetic term and utilize it as a difference in the electron-electron term.  This Hamiltonian is shown in Eq. (\ref{eq:new}).  It's energies completely agree with those of the the radial and cylindrical cases, and by allowing this symmetry breaking in the x-coordinate can be use for the potential discussed in Section II $\S$E.

This Planar Hamiltonian, $\mathcal{H}_{P}$:
\begin{equation}
\label{eq:new}
\begin{split}
\mathcal{H}_{P}=&\frac{1}{2}\sum\limits^{N}_{i=1}\frac{1}{y_{i}^{2}}+\sum\limits_{i=1}^{N}V(x_{i}, y_{i}, z_{i})\\
&+\sum\limits_{i=1}^{N}\sum\limits_{j=i+1}^{N-1}\frac{1}{\sqrt{(z_{i}\!\!-\!\!z_{j})^{2}\!\!+\!\!(x_{i}\!\!-\!\!x_{j})^{2}\!\!+\!\!y^{2}_{i}\!\!+\!\!y^{2}_{j}}}
\end{split}
\end{equation}
was found to allow -but not require- the breaking of symmetry, as it was applied to the previously discussed linear systems and had extremely good agreement, shown in Figure \ref{fig:Global}.  The relation between the Cartesian coordinates used in the above and the geometry of the system is shown in Figure \ref{fig:coords}  When applied to our system, containing the corrected trajectory, this new Hamiltonian performs both smoothing with smoothly evolving trajectories- as seen in Figure \ref{fig:Local}.  This planar Dimensional Scaled Hamiltonian was capable of adequate description of the system in all three of the regimes discussed earlier.

\section{Results and Discussion}

We shall concern ourselves with a discussion of the binding energies (B.E.) for the following species: H$^{-}$, H$^{-2}$, He$^{-}$, He$^{-2}$, He$^{-3}$; where the B.E. is the negative of the detachment energy for a single 'excess' electron shown in Eq. (\ref{eq:bedef}), where N signifies the number of electrons for a species.
\begin{equation}
\label{eq:bedef}
B.E.=E_{H}(N)-E_{H}(N-1)
\end{equation}
  Figure \ref{fig:BEs} Displays the binding energies for the two species, Z=1 and Z=2.  From this figure we can see a clear maximum binding energy for the H$^{-}$ species (top left) at, roughly, $\alpha_{0}$ = 10; this energy shows a stability of the second electron of 0.047 Hartree (1.28 eV).  Also shown in Figure \ref{fig:BEs} the B.E. curve for H$^{2-}$ (top right), showing a stable binding of 0.00012 Hartree (0.0033eV).  This, the B.E. of H$^{2-}$, reaches it's minimal value asymptotically with increasing $\alpha_{0}$ implying the addition of any further electrons will not be allowed; this fact was verified by performing the requisite minimization, and if the mass gauge was not applied to the system the number of allowed additional electrons would increase unrealistically and seemingly without bound as the laser intensity is increased.

The middle, left plot in Figure \ref{fig:BEs} shows the binding of a third electron to Helium at $\alpha_{0}$ equals, roughly, 10 again; the binding energy for this species at it's greatest magnitude is extrapolated to be 0.057 Hartree (1.55 eV).  The second 'additional' electron to Helium (middle right) is most stable at $\alpha_{0}$=90, with a B.E. of 0.007 Hartree (0.19 eV).  The fifth and final electron which can stable bind to a Z=2 center (bottom center) is, similarly to the H$^{2-}$ species, a terminal binding who reaches an asymptotic stability with increasing alpha, the B.E. is 0.0004 Hartree (0.011 eV).

Of the two relativistic corrections accounted for in the above framework, the trajectory is the paramount addition. An examination of Figures \ref{fig:Local} and \ref{fig:Global} shall be required for the subsequent discussions. The introduction of the planar Large-D Hamiltonian, $\mathcal{H}_{P}$, for the systems was a boom which aligns itself in agreement with the previous works -see Figure \ref{fig:Local} a and compare plots a and c in Figure \ref{fig:Global}- and yet out preforms the previous equation when this relativistic trajectory is introduced, to see this compared plots b and c in Figure \ref{fig:Global}.

By comparing plots a and d of Figure \ref{fig:Global} it can be seen that the raw energies and the binding energies between the non-relativistic trajectory (plotted with $\mathcal{H}_{DA}$) and the relativistic trajectory are not extreme. This can be verified and more clearly seen by examination of Figure \ref{fig:Local} c, wherein both the non-relativistic and relativistic binding energies are shown; although the energies become quite different with increasing $\alpha_{0}$, the field intensity which yields the most stable binding energy is same and the most stable binding energy deviates only in the thousandth of a Hartree. 
Consideration of the mass gauge for this system provides a very slight correction within the values of $\alpha_{0}$ examined here; where although these values of $\alpha_{0}$ indicate laser field strengths on the order of atomic units and greater (in competition with the Coulomb potential of the center) they are in no way strong enough to generate quivering electron masses, m$_{r}$, which deviate significantly from the invariant mass, m$_{0}$. Due to this, the deviation of the binding energies due to consideration of $m_{r}$ over $m_{0}$ is also slightly less than breath-taking, this can be seen in Figure \ref{fig:Local}b and again impressed by examination of the deviation of the raw energies of Hydrogen in Figure \ref{fig:Local}d. To be gleaned from this is that within the examined field strengths, $m_{r}$ deviates very little from $m_{0}$, but more significant is the fact that the overall consideration of relativistic effects does destabilize the system, but not by an appreciable amount.

Figure \ref{fig:probconts} helps emphasis the differences between the corrected and non-corrected systems.  The top level of the figure is the probability density, $ | \psi |  ^{2} $, of corrected (left) and non-corrected (right), the density in the outer orbital centers is lower due to a more diffuse spread of probability density at these locations compared to the non-corrected.  The mid-level of the plot is the negative probability density, superimposed on the probability density function is the trajectory of the system which it describes.  Below all else is a contour plot of each system, again emphasis is merited on the more diffuse spread in the corrected system due to the evolution of the parametric trajectory as opposed to the linear oscillating trajectory of the non-corrected system.

Figure \ref{fig:BEs3D} shows a plot of the B.E. For the H$^{-}$ (left) and He$^{-}$ (right) systems from both the SCF (blue) and D-Scaled (red) methods. The lower B.E. plots in Fig. \ref{fig:BEs3D}  have been normalized to the minimum value to show the tight agreement between the qualitative assessment of the scaling procedure and the SCF method. The Dimensionally Scaled minimization problem bore 'exact' position of the electrons as $(x_{1}, y_{1}, z_{1}; \alpha_{0})$=$(4.1660\times10^{-10}, 5.4461, -12.2387; 20)$ for the Z=1, N=1 system and $(x_{1}, y_{1}, z_{1}; x_{2}, y_{2}, z_{2}; \alpha_{0})$=$(6.4472\times10^{-8}, 5.0386, 16.5852; -6.4472\times10^{-8}, 5.0386, -16.5852; 20)$  for the Z=1, N=2 system; these localized electron positions are similarly predictable as they attempt to bind to regoins where the angular velocity of the nucleus is lowest in the KH Frame; these locations would most notable be the $\frac{\pi}{4}$'s the trajectory.  With a single electron the central point of the parametric curve, set a origin in our calculations, binds the electron strongly; as more electrons are introduced they are situated at locations which minimize the electron-electron repulsion of the system.   A Mulliken population analysis of the systems shows that the orbitals about the central charge typically possess a smaller number of electrons, except in the N=3 (and assumably the N=5 case); the results of such population assessments can be seen in Table \ref{my_table}. In this way we are able to verify not only the energetic behavior of the D-Scaled Hamiltonian but its treatment of the electrons in space.
\section{Conclusion}
\label{Conclu}
It has been shown that even under conditions of relativity multiply-charge atomic ions should still be achievable within the confines of an intense laser field. The stability of several atomic-anions has been found and discussed, some ions are on the order of an entire electron volt more stable than the ionized system. The importance of the general dimensional scaling procedure was verified not only via the energetics, but with a comparison of the 'exact' locations of the electrons as predicted by Dimensional Scaling as they compare to the probability densities from the standard SCF procedure. These species were found to be stable and should, therefore, be experimentally realizable.  Stability of simple molecular systems in super-intense laser fields have been previous discussed here\cite{dichotomy,0953-4075-37-15-010,0953-4075-35-17-103}.  This dimensionally scaled framework with relativistic corrections yields itself easily to a description of molecules and molecular ions within the confines of super-intense laser fields, which shall be undertaken next.

\section{Aknowledgments}
We would like to thank the Army Research Offices for funding this project.  Jiaxiang Wang would like to thank NSF-China for the support by Grant No. 10974056.  We would also like to thank Prof. Dudley Herschbach of stimulating discussion of the materials within this paper.

\section{Appendix}
\subsection{Elliptical and Circular Polarizations}
Within the above concern was only given to the relativistically corrected linear polarized light, this is because similar corrections placed on elliptical polarized light yield no new -or interesting- phenomenon. Following similar mathematics as to achieve an analogous relativisticly considered trajectory would consider the electronic coupling as before and consider a minor weighted coupling to the magnetic field, as in the linear case before. An elliptically polarized laser fired in the y-direction with the electronic major (minor) axis oriented in the x(z)-direction yields the trajectories
\begin{equation}
\vec \alpha_{\vec E, Elliptical}(t) = \{\epsilon_{1}cos(\phi), 0, \epsilon_{2}sin(\phi) \} \label{EllE}
\end{equation}
\begin{equation}
\vec \alpha_{\vec B, Elliptical}(t) = \{ - \beta_{2}sin(\phi), 0, \beta_{1}cos(\phi) \}. \label{EllB}
\end{equation}
Within Eqs. (\ref{EllE}-\ref{EllB}) the amplitude in each the major and minor axis is denoted by the subscript 1 and 2, respectively, and the $\epsilon$ and $\beta$ are the coefficients of the electronic and magnetic components. The trajectory generated by both the above biases applied to free particle merely generates a new ellipse with a major and minor amplitude mediated between those of the above and tilted by an angle respective to the coefficients.
\begin{equation}
\vec \alpha = \{ \epsilon_{1}cos(t) - \beta_{2}sin(t), 0, \epsilon_{2}sin(\phi)+ \beta_{1}cos(\phi) \}.
\end{equation}
This can be seen graphically in Fig. \ref{fig:SeperateEll} which displays the individual electronic and magnetic trajectories and then in Fig. \ref{fig:combinedEll} which displays the combined trajectory which would be followed by a free particle traveling within this electromagnetic field.

If this scheme is applied to a circularly polarized field the same mathematics will appear but the coefficients within the trajectories will be reduced to $\epsilon_{1}=\epsilon_{2}=\epsilon$ and $\beta_{1}=\beta_{2}=\beta$. This will yield the same uninteresting phenomenon, but merely present it as the mediate of two circles with no change in the orientation angle as there is no unique point of reference on a circle.

\newpage

\bibliographystyle{apsrev4-1}
\bibliography{RelaCorsDScal}

\begin{thebibliography}{36}%
\makeatletter
\providecommand \@ifxundefined [1]{%
 \@ifx{#1\undefined}
}%
\providecommand \@ifnum [1]{%
 \ifnum #1\expandafter \@firstoftwo
 \else \expandafter \@secondoftwo
 \fi
}%
\providecommand \@ifx [1]{%
 \ifx #1\expandafter \@firstoftwo
 \else \expandafter \@secondoftwo
 \fi
}%
\providecommand \natexlab [1]{#1}%
\providecommand \enquote  [1]{``#1''}%
\providecommand \bibnamefont  [1]{#1}%
\providecommand \bibfnamefont [1]{#1}%
\providecommand \citenamefont [1]{#1}%
\providecommand \href@noop [0]{\@secondoftwo}%
\providecommand \href [0]{\begingroup \@sanitize@url \@href}%
\providecommand \@href[1]{\@@startlink{#1}\@@href}%
\providecommand \@@href[1]{\endgroup#1\@@endlink}%
\providecommand \@sanitize@url [0]{\catcode `\\12\catcode `\$12\catcode
  `\&12\catcode `\#12\catcode `\^12\catcode `\_12\catcode `\%12\relax}%
\providecommand \@@startlink[1]{}%
\providecommand \@@endlink[0]{}%
\providecommand \url  [0]{\begingroup\@sanitize@url \@url }%
\providecommand \@url [1]{\endgroup\@href {#1}{\urlprefix }}%
\providecommand \urlprefix  [0]{URL }%
\providecommand \Eprint [0]{\href }%
\providecommand \doibase [0]{http://dx.doi.org/}%
\providecommand \selectlanguage [0]{\@gobble}%
\providecommand \bibinfo  [0]{\@secondoftwo}%
\providecommand \bibfield  [0]{\@secondoftwo}%
\providecommand \translation [1]{[#1]}%
\providecommand \BibitemOpen [0]{}%
\providecommand \bibitemStop [0]{}%
\providecommand \bibitemNoStop [0]{.\EOS\space}%
\providecommand \EOS [0]{\spacefactor3000\relax}%
\providecommand \BibitemShut  [1]{\csname bibitem#1\endcsname}%
\let\auto@bib@innerbib\@empty
\bibitem [{\citenamefont {Gavrila}(2002)}]{0953-4075-35-18-201}%
  \BibitemOpen
  \bibfield  {author} {\bibinfo {author} {\bibfnamefont {M.}~\bibnamefont
  {Gavrila}},\ }\href {http://stacks.iop.org/0953-4075/35/i=18/a=201}
  {\bibfield  {journal} {\bibinfo  {journal} {J. Phys. B}\ }\textbf {\bibinfo
  {volume} {35}},\ \bibinfo {pages} {R147} (\bibinfo {year}
  {2002})}\BibitemShut {NoStop}%
\bibitem [{\citenamefont {Gavrila}(1992)}]{Gavrila_Book}%
  \BibitemOpen
  \bibfield  {author} {\bibinfo {author} {\bibfnamefont {M.}~\bibnamefont
  {Gavrila}},\ }\href@noop {} {\emph {\bibinfo {title} {Atoms in Intense Laser
  Fields}}}\ (\bibinfo  {publisher} {Academic Press, NY},\ \bibinfo {year}
  {1992})\BibitemShut {NoStop}%
\bibitem [{\citenamefont {Eberly}\ and\ \citenamefont
  {Kulander}(1993)}]{eberly}%
  \BibitemOpen
  \bibfield  {author} {\bibinfo {author} {\bibfnamefont {J.~H.}\ \bibnamefont
  {Eberly}}\ and\ \bibinfo {author} {\bibfnamefont {K.~C.}\ \bibnamefont
  {Kulander}},\ }\href {\doibase 10.1126/science.262.5137.1229} {\bibfield
  {journal} {\bibinfo  {journal} {Science}\ }\textbf {\bibinfo {volume}
  {262}},\ \bibinfo {pages} {1229} (\bibinfo {year} {1993})}\BibitemShut
  {NoStop}%
\bibitem [{\citenamefont {{Pegg}}\ \emph {et~al.}(1987)\citenamefont {{Pegg}},
  \citenamefont {{Thompson}}, \citenamefont {{Compton}},\ and\ \citenamefont
  {{Alton}}}]{1987PhRvL..59.2267P}%
  \BibitemOpen
  \bibfield  {author} {\bibinfo {author} {\bibfnamefont {D.~J.}\ \bibnamefont
  {{Pegg}}}, \bibinfo {author} {\bibfnamefont {J.~S.}\ \bibnamefont
  {{Thompson}}}, \bibinfo {author} {\bibfnamefont {R.~N.}\ \bibnamefont
  {{Compton}}}, \ and\ \bibinfo {author} {\bibfnamefont {G.~D.}\ \bibnamefont
  {{Alton}}},\ }\href {\doibase 10.1103/PhysRevLett.59.2267} {\bibfield
  {journal} {\bibinfo  {journal} {Phys. Rev. Lett.}\ }\textbf {\bibinfo
  {volume} {59}},\ \bibinfo {pages} {2267} (\bibinfo {year}
  {1987})}\BibitemShut {NoStop}%
\bibitem [{\citenamefont {{Berkovits}}\ \emph {et~al.}(1995)\citenamefont
  {{Berkovits}}, \citenamefont {{Boaretto}}, \citenamefont {{Ghelberg}},
  \citenamefont {{Heber}},\ and\ \citenamefont {{Paul}}}]{1995PhRvL..75..414B}%
  \BibitemOpen
  \bibfield  {author} {\bibinfo {author} {\bibfnamefont {D.}~\bibnamefont
  {{Berkovits}}}, \bibinfo {author} {\bibfnamefont {E.}~\bibnamefont
  {{Boaretto}}}, \bibinfo {author} {\bibfnamefont {S.}~\bibnamefont
  {{Ghelberg}}}, \bibinfo {author} {\bibfnamefont {O.}~\bibnamefont {{Heber}}},
  \ and\ \bibinfo {author} {\bibfnamefont {M.}~\bibnamefont {{Paul}}},\ }\href
  {\doibase 10.1103/PhysRevLett.75.414} {\bibfield  {journal} {\bibinfo
  {journal} {Phys. Rev. Lett.}\ }\textbf {\bibinfo {volume} {75}},\ \bibinfo
  {pages} {414} (\bibinfo {year} {1995})}\BibitemShut {NoStop}%
\bibitem [{\citenamefont {Scheller}\ \emph {et~al.}(1995)\citenamefont
  {Scheller}, \citenamefont {Compton},\ and\ \citenamefont
  {Cederbaum}}]{Scheller17111995}%
  \BibitemOpen
  \bibfield  {author} {\bibinfo {author} {\bibfnamefont {M.~K.}\ \bibnamefont
  {Scheller}}, \bibinfo {author} {\bibfnamefont {R.~N.}\ \bibnamefont
  {Compton}}, \ and\ \bibinfo {author} {\bibfnamefont {L.~S.}\ \bibnamefont
  {Cederbaum}},\ }\href {\doibase 10.1126/science.270.5239.1160} {\bibfield
  {journal} {\bibinfo  {journal} {Science}\ }\textbf {\bibinfo {volume}
  {270}},\ \bibinfo {pages} {1160} (\bibinfo {year} {1995})}\BibitemShut
  {NoStop}%
\bibitem [{\citenamefont {Simons}(2011)}]{SimonsII}%
  \BibitemOpen
  \bibfield  {author} {\bibinfo {author} {\bibfnamefont {J.}~\bibnamefont
  {Simons}},\ }\href {\doibase 10.1146/annurev-physchem-032210-103547}
  {\bibfield  {journal} {\bibinfo  {journal} {Ann. Rev. Phys. Chem.}\ }\textbf
  {\bibinfo {volume} {62}},\ \bibinfo {pages} {107} (\bibinfo {year}
  {2011})}\BibitemShut {NoStop}%
\bibitem [{\citenamefont {Simons}(2008)}]{SimonsI}%
  \BibitemOpen
  \bibfield  {author} {\bibinfo {author} {\bibfnamefont {J.}~\bibnamefont
  {Simons}},\ }\href {\doibase 10.1021/jp711490b} {\bibfield  {journal}
  {\bibinfo  {journal} {J. Phys. B}\ }\textbf {\bibinfo {volume} {112}},\
  \bibinfo {pages} {6401} (\bibinfo {year} {2008})}\BibitemShut {NoStop}%
\bibitem [{\citenamefont {{Lieb}}(1984)}]{1984PhRvA..29.3018L}%
  \BibitemOpen
  \bibfield  {author} {\bibinfo {author} {\bibfnamefont {E.~H.}\ \bibnamefont
  {{Lieb}}},\ }\href {\doibase 10.1103/PhysRevA.29.3018} {\bibfield  {journal}
  {\bibinfo  {journal} {Phys. Rev. A}\ }\textbf {\bibinfo {volume} {29}},\
  \bibinfo {pages} {3018} (\bibinfo {year} {1984})}\BibitemShut {NoStop}%
\bibitem [{\citenamefont {{Chang}}\ \emph {et~al.}(1987)\citenamefont
  {{Chang}}, \citenamefont {{McKeown}}, \citenamefont {{Milner}},\ and\
  \citenamefont {{Labrenz}}}]{1987PhRvA..35.3949C}%
  \BibitemOpen
  \bibfield  {author} {\bibinfo {author} {\bibfnamefont {K.~H.}\ \bibnamefont
  {{Chang}}}, \bibinfo {author} {\bibfnamefont {R.~D.}\ \bibnamefont
  {{McKeown}}}, \bibinfo {author} {\bibfnamefont {R.~G.}\ \bibnamefont
  {{Milner}}}, \ and\ \bibinfo {author} {\bibfnamefont {J.}~\bibnamefont
  {{Labrenz}}},\ }\href {\doibase 10.1103/PhysRevA.35.3949} {\bibfield
  {journal} {\bibinfo  {journal} {Phys. Rev. A}\ }\textbf {\bibinfo {volume}
  {35}},\ \bibinfo {pages} {3949} (\bibinfo {year} {1987})}\BibitemShut
  {NoStop}%
\bibitem [{\citenamefont {Bates}(1990)}]{Bates19901}%
  \BibitemOpen
  \bibfield  {author} {\bibinfo {author} {\bibfnamefont {D.~R.}\ \bibnamefont
  {Bates}},\ }in\ \href {\doibase DOI: 10.1016/S1049-250X(08)60148-2} {\emph
  {\bibinfo {booktitle} {Negative Ions: Structure and Spectra}}},\ \bibinfo
  {series} {Advances In Atomic, Molecular, and Optical Physics}, Vol.~\bibinfo
  {volume} {27},\ \bibinfo {editor} {edited by\ \bibinfo {editor}
  {\bibfnamefont {S.~D.}\ \bibnamefont {Bates}}\ and\ \bibinfo {editor}
  {\bibfnamefont {B.}~\bibnamefont {Bederson}}}\ (\bibinfo  {publisher}
  {Academic Press},\ \bibinfo {year} {1990})\ pp.\ \bibinfo {pages} {1 --
  80}\BibitemShut {NoStop}%
\bibitem [{\citenamefont {Sergeev}\ and\ \citenamefont
  {Kais}(1999)}]{0305-4470-32-39-312}%
  \BibitemOpen
  \bibfield  {author} {\bibinfo {author} {\bibfnamefont {A.~V.}\ \bibnamefont
  {Sergeev}}\ and\ \bibinfo {author} {\bibfnamefont {S.}~\bibnamefont {Kais}},\
  }\href {http://stacks.iop.org/0305-4470/32/i=39/a=312} {\bibfield  {journal}
  {\bibinfo  {journal} {J. Phys. A}\ }\textbf {\bibinfo {volume} {32}},\
  \bibinfo {pages} {6891} (\bibinfo {year} {1999})}\BibitemShut {NoStop}%
\bibitem [{\citenamefont {Serra}\ and\ \citenamefont
  {Kais}(2003)}]{Serra2003205}%
  \BibitemOpen
  \bibfield  {author} {\bibinfo {author} {\bibfnamefont {P.}~\bibnamefont
  {Serra}}\ and\ \bibinfo {author} {\bibfnamefont {S.}~\bibnamefont {Kais}},\
  }\href {\doibase DOI: 10.1016/S0009-2614(03)00371-3} {\bibfield  {journal}
  {\bibinfo  {journal} {Chem. Phys. Lett.}\ }\textbf {\bibinfo {volume}
  {372}},\ \bibinfo {pages} {205 } (\bibinfo {year} {2003})}\BibitemShut
  {NoStop}%
\bibitem [{\citenamefont {{Wei}}\ \emph {et~al.}(2007)\citenamefont {{Wei}},
  \citenamefont {{Kais}},\ and\ \citenamefont
  {{Herschbach}}}]{2007JChPh.127i4301W}%
  \BibitemOpen
  \bibfield  {author} {\bibinfo {author} {\bibfnamefont {Q.}~\bibnamefont
  {{Wei}}}, \bibinfo {author} {\bibfnamefont {S.}~\bibnamefont {{Kais}}}, \
  and\ \bibinfo {author} {\bibfnamefont {D.}~\bibnamefont {{Herschbach}}},\
  }\href {\doibase 10.1063/1.2768037} {\bibfield  {journal} {\bibinfo
  {journal} {J. Chem. Phys.}\ }\textbf {\bibinfo {volume} {127}},\ \bibinfo
  {pages} {094301} (\bibinfo {year} {2007})}\BibitemShut {NoStop}%
\bibitem [{\citenamefont {{Wei}}\ \emph {et~al.}(2006)\citenamefont {{Wei}},
  \citenamefont {{Kais}},\ and\ \citenamefont
  {{Moiseyev}}}]{2006JChPh.124t1108W}%
  \BibitemOpen
  \bibfield  {author} {\bibinfo {author} {\bibfnamefont {Q.}~\bibnamefont
  {{Wei}}}, \bibinfo {author} {\bibfnamefont {S.}~\bibnamefont {{Kais}}}, \
  and\ \bibinfo {author} {\bibfnamefont {N.}~\bibnamefont {{Moiseyev}}},\
  }\href {\doibase 10.1063/1.2207619} {\bibfield  {journal} {\bibinfo
  {journal} {J. Chem. Phys.}\ }\textbf {\bibinfo {volume} {124}},\ \bibinfo
  {pages} {201108} (\bibinfo {year} {2006})}\BibitemShut {NoStop}%
\bibitem [{\citenamefont {Pont}\ \emph {et~al.}(1988)\citenamefont {Pont},
  \citenamefont {Walet}, \citenamefont {Gavrila},\ and\ \citenamefont
  {McCurdy}}]{PhysRevLett.61.939}%
  \BibitemOpen
  \bibfield  {author} {\bibinfo {author} {\bibfnamefont {M.}~\bibnamefont
  {Pont}}, \bibinfo {author} {\bibfnamefont {N.~R.}\ \bibnamefont {Walet}},
  \bibinfo {author} {\bibfnamefont {M.}~\bibnamefont {Gavrila}}, \ and\
  \bibinfo {author} {\bibfnamefont {C.~W.}\ \bibnamefont {McCurdy}},\ }\href
  {\doibase 10.1103/PhysRevLett.61.939} {\bibfield  {journal} {\bibinfo
  {journal} {Phys. Rev. Lett.}\ }\textbf {\bibinfo {volume} {61}},\ \bibinfo
  {pages} {939} (\bibinfo {year} {1988})}\BibitemShut {NoStop}%
\bibitem [{\citenamefont {van Duijn}\ \emph {et~al.}(1996)\citenamefont {van
  Duijn}, \citenamefont {Gavrila},\ and\ \citenamefont {Muller}}]{vanDuijin}%
  \BibitemOpen
  \bibfield  {author} {\bibinfo {author} {\bibfnamefont {E.}~\bibnamefont {van
  Duijn}}, \bibinfo {author} {\bibfnamefont {M.}~\bibnamefont {Gavrila}}, \
  and\ \bibinfo {author} {\bibfnamefont {H.~G.}\ \bibnamefont {Muller}},\
  }\href {http://prl.aps.org/pdf/PRL/v77/i18/p3759_1} {\bibfield  {journal}
  {\bibinfo  {journal} {Phys. Rev. Lett.}\ }\textbf {\bibinfo {volume} {77}},\
  \bibinfo {pages} {3759} (\bibinfo {year} {1996})}\BibitemShut {NoStop}%
\bibitem [{\citenamefont {Wei}\ \emph {et~al.}(2007)\citenamefont {Wei},
  \citenamefont {Kais},\ and\ \citenamefont {Moiseyev}}]{PhysRevA.76.013407}%
  \BibitemOpen
  \bibfield  {author} {\bibinfo {author} {\bibfnamefont {Q.}~\bibnamefont
  {Wei}}, \bibinfo {author} {\bibfnamefont {S.}~\bibnamefont {Kais}}, \ and\
  \bibinfo {author} {\bibfnamefont {N.}~\bibnamefont {Moiseyev}},\ }\href
  {\doibase 10.1103/PhysRevA.76.013407} {\bibfield  {journal} {\bibinfo
  {journal} {Phys. Rev. A}\ }\textbf {\bibinfo {volume} {76}},\ \bibinfo
  {pages} {013407} (\bibinfo {year} {2007})}\BibitemShut {NoStop}%
\bibitem [{\citenamefont {{Wei}}\ \emph {et~al.}(2008)\citenamefont {{Wei}},
  \citenamefont {{Kais}},\ and\ \citenamefont
  {{Herschbach}}}]{2008JChPh.129u4110W}%
  \BibitemOpen
  \bibfield  {author} {\bibinfo {author} {\bibfnamefont {Q.}~\bibnamefont
  {{Wei}}}, \bibinfo {author} {\bibfnamefont {S.}~\bibnamefont {{Kais}}}, \
  and\ \bibinfo {author} {\bibfnamefont {D.}~\bibnamefont {{Herschbach}}},\
  }\href {\doibase 10.1063/1.3027451} {\bibfield  {journal} {\bibinfo
  {journal} {J. of Chem. Phys.}\ }\textbf {\bibinfo {volume} {129}},\ \bibinfo
  {pages} {214110} (\bibinfo {year} {2008})}\BibitemShut {NoStop}%
\bibitem [{\citenamefont {Joachain}\ \emph {et~al.}(2000)\citenamefont
  {Joachain}, \citenamefont {Dorr},\ and\ \citenamefont
  {Kylstra}}]{Joachain2000225}%
  \BibitemOpen
  \bibfield  {author} {\bibinfo {author} {\bibfnamefont {C.}~\bibnamefont
  {Joachain}}, \bibinfo {author} {\bibfnamefont {M.}~\bibnamefont {Dorr}}, \
  and\ \bibinfo {author} {\bibfnamefont {N.}~\bibnamefont {Kylstra}},\ }in\
  \href {\doibase DOI: 10.1016/S1049-250X(08)60188-3} {\emph {\bibinfo
  {booktitle} {High-Intensity Laser-Atom Physics}}},\ \bibinfo {series}
  {Advances In Atomic, Molecular, and Optical Physics}, Vol.~\bibinfo {volume}
  {42},\ \bibinfo {editor} {edited by\ \bibinfo {editor} {\bibfnamefont
  {B.}~\bibnamefont {Bederson}}\ and\ \bibinfo {editor} {\bibfnamefont
  {H.}~\bibnamefont {Walther}}}\ (\bibinfo  {publisher} {Academic Press},\
  \bibinfo {year} {2000})\ pp.\ \bibinfo {pages} {225 -- 286}\BibitemShut
  {NoStop}%
\bibitem [{\citenamefont {{Joachain}}\ and\ \citenamefont
  {{Kylstra}}(2003)}]{2003PhyS...68C..72J}%
  \BibitemOpen
  \bibfield  {author} {\bibinfo {author} {\bibfnamefont {C.~J.}\ \bibnamefont
  {{Joachain}}}\ and\ \bibinfo {author} {\bibfnamefont {N.~J.}\ \bibnamefont
  {{Kylstra}}},\ }\href {\doibase 10.1238/Physica.Regular.068aC0072} {\bibfield
   {journal} {\bibinfo  {journal} {Physica Scripta}\ }\textbf {\bibinfo
  {volume} {68}},\ \bibinfo {pages} {C72+} (\bibinfo {year}
  {2003})}\BibitemShut {NoStop}%
\bibitem [{\citenamefont {Brown}\ and\ \citenamefont
  {Kibble}(1964)}]{PhysRev.133.A705}%
  \BibitemOpen
  \bibfield  {author} {\bibinfo {author} {\bibfnamefont {L.~S.}\ \bibnamefont
  {Brown}}\ and\ \bibinfo {author} {\bibfnamefont {T.~W.~B.}\ \bibnamefont
  {Kibble}},\ }\href {\doibase 10.1103/PhysRev.133.A705} {\bibfield  {journal}
  {\bibinfo  {journal} {Phys. Rev.}\ }\textbf {\bibinfo {volume} {133}},\
  \bibinfo {pages} {A705} (\bibinfo {year} {1964})}\BibitemShut {NoStop}%
\bibitem [{\citenamefont {Eberly}\ and\ \citenamefont
  {Sleeper}(1968)}]{PhysRev.176.1570}%
  \BibitemOpen
  \bibfield  {author} {\bibinfo {author} {\bibfnamefont {J.~H.}\ \bibnamefont
  {Eberly}}\ and\ \bibinfo {author} {\bibfnamefont {A.}~\bibnamefont
  {Sleeper}},\ }\href {\doibase 10.1103/PhysRev.176.1570} {\bibfield  {journal}
  {\bibinfo  {journal} {Phys. Rev.}\ }\textbf {\bibinfo {volume} {176}},\
  \bibinfo {pages} {1570} (\bibinfo {year} {1968})}\BibitemShut {NoStop}%
\bibitem [{\citenamefont {McMurchie}\ and\ \citenamefont
  {Davidson}(1978)}]{McMurchie1978218}%
  \BibitemOpen
  \bibfield  {author} {\bibinfo {author} {\bibfnamefont {L.~E.}\ \bibnamefont
  {McMurchie}}\ and\ \bibinfo {author} {\bibfnamefont {E.~R.}\ \bibnamefont
  {Davidson}},\ }\href {\doibase DOI: 10.1016/0021-9991(78)90092-X} {\bibfield
  {journal} {\bibinfo  {journal} {J. Comp. Phys.}\ }\textbf {\bibinfo {volume}
  {26}},\ \bibinfo {pages} {218 } (\bibinfo {year} {1978})}\BibitemShut
  {NoStop}%
\bibitem [{\citenamefont {Szabo}\ and\ \citenamefont
  {Ostlund}(1996)}]{Szabo_Ostlund_1996}%
  \BibitemOpen
  \bibfield  {author} {\bibinfo {author} {\bibfnamefont {A.}~\bibnamefont
  {Szabo}}\ and\ \bibinfo {author} {\bibfnamefont {N.~S.}\ \bibnamefont
  {Ostlund}},\ }\href {http://books.google.com/books?id=6mV9gYzEkgIC} {\emph
  {\bibinfo {title} {Modern quantum chemistry: introduction to advanced
  electronic structure theory}}}\ (\bibinfo  {publisher} {Dover Publications},\
  \bibinfo {year} {1996})\ p.\ \bibinfo {pages} {480}\BibitemShut {NoStop}%
\bibitem [{\citenamefont {{Boys}}(1950)}]{1950RSPSA.200..542B}%
  \BibitemOpen
  \bibfield  {author} {\bibinfo {author} {\bibfnamefont {S.~F.}\ \bibnamefont
  {{Boys}}},\ }\href {\doibase 10.1098/rspa.1950.0036} {\bibfield  {journal}
  {\bibinfo  {journal} {Royal Society of London Proceedings Series A}\ }\textbf
  {\bibinfo {volume} {200}},\ \bibinfo {pages} {542} (\bibinfo {year}
  {1950})}\BibitemShut {NoStop}%
\bibitem [{\citenamefont {Herschbach}(1986)}]{DudsPap}%
  \BibitemOpen
  \bibfield  {author} {\bibinfo {author} {\bibfnamefont {D.~R.}\ \bibnamefont
  {Herschbach}},\ }\href {\doibase DOI:10.1063/1.450584} {\bibfield  {journal}
  {\bibinfo  {journal} {J. Chem. Phys.}\ }\textbf {\bibinfo {volume} {84}},\
  \bibinfo {pages} {838} (\bibinfo {year} {1986})}\BibitemShut {NoStop}%
\bibitem [{\citenamefont {Herschbach}\ \emph {et~al.}(1993)\citenamefont
  {Herschbach}, \citenamefont {Aver},\ and\ \citenamefont
  {Goscinkis}}]{Herschbach}%
  \BibitemOpen
  \bibfield  {author} {\bibinfo {author} {\bibfnamefont {D.}~\bibnamefont
  {Herschbach}}, \bibinfo {author} {\bibfnamefont {J.}~\bibnamefont {Aver}}, \
  and\ \bibinfo {author} {\bibfnamefont {O.}~\bibnamefont {Goscinkis}},\
  }\href@noop {} {\emph {\bibinfo {title} {Dimensional Scaling in Chemical
  Physics}}}\ (\bibinfo  {publisher} {Kluwer Academic Publishers},\ \bibinfo
  {year} {1993})\ p.\ \bibinfo {pages} {480}\BibitemShut {NoStop}%
\bibitem [{\citenamefont {Svidzinsky}\ \emph {et~al.}(2008)\citenamefont
  {Svidzinsky}, \citenamefont {Chen}, \citenamefont {Chin},\ and\ \citenamefont
  {et~al.}}]{Bohrs}%
  \BibitemOpen
  \bibfield  {author} {\bibinfo {author} {\bibfnamefont {A.}~\bibnamefont
  {Svidzinsky}}, \bibinfo {author} {\bibfnamefont {G.}~\bibnamefont {Chen}},
  \bibinfo {author} {\bibfnamefont {S.}~\bibnamefont {Chin}}, \ and\ \bibinfo
  {author} {\bibnamefont {et~al.}},\ }\href {\doibase
  10.1080/01442350802364664} {\bibfield  {journal} {\bibinfo  {journal} {Int.
  Rev. Phys. Chem.}\ }\textbf {\bibinfo {volume} {27}},\ \bibinfo {pages} {665}
  (\bibinfo {year} {2008})}\BibitemShut {NoStop}%
\bibitem [{\citenamefont {Kais}\ and\ \citenamefont
  {Herschbach}(1993)}]{sabre}%
  \BibitemOpen
  \bibfield  {author} {\bibinfo {author} {\bibfnamefont {S.}~\bibnamefont
  {Kais}}\ and\ \bibinfo {author} {\bibfnamefont {D.~R.}\ \bibnamefont
  {Herschbach}},\ }\href {http://link.aip.org/link/doi/10.1063/1.466319}
  {\bibfield  {journal} {\bibinfo  {journal} {J. Chem. Phys.}\ }\textbf
  {\bibinfo {volume} {100}},\ \bibinfo {pages} {4367} (\bibinfo {year}
  {1993})}\BibitemShut {NoStop}%
\bibitem [{\citenamefont {Dunn}\ \emph {et~al.}(1994)\citenamefont {Dunn},
  \citenamefont {Germann}, \citenamefont {Goodson}, \citenamefont {Traynor},
  \citenamefont {Morgan}, \citenamefont {Watson},\ and\ \citenamefont
  {Herschbach}}]{Dunn}%
  \BibitemOpen
  \bibfield  {author} {\bibinfo {author} {\bibfnamefont {M.}~\bibnamefont
  {Dunn}}, \bibinfo {author} {\bibfnamefont {T.~C.}\ \bibnamefont {Germann}},
  \bibinfo {author} {\bibfnamefont {D.~Z.}\ \bibnamefont {Goodson}}, \bibinfo
  {author} {\bibfnamefont {C.~A.}\ \bibnamefont {Traynor}}, \bibinfo {author}
  {\bibfnamefont {J.~D.}\ \bibnamefont {Morgan}}, \bibinfo {author}
  {\bibfnamefont {D.~K.}\ \bibnamefont {Watson}}, \ and\ \bibinfo {author}
  {\bibfnamefont {D.~R.}\ \bibnamefont {Herschbach}},\ }\href
  {http://link.aip.org/link/doi/10.1063/1.467314} {\bibfield  {journal}
  {\bibinfo  {journal} {J. Chem. Phys.}\ }\textbf {\bibinfo {volume} {101}},\
  \bibinfo {pages} {5987} (\bibinfo {year} {1994})}\BibitemShut {NoStop}%
\bibitem [{\citenamefont {Avery}\ \emph {et~al.}(1991)\citenamefont {Avery},
  \citenamefont {Goodson},\ and\ \citenamefont {Herschabach}}]{Avery}%
  \BibitemOpen
  \bibfield  {author} {\bibinfo {author} {\bibfnamefont {J.}~\bibnamefont
  {Avery}}, \bibinfo {author} {\bibfnamefont {D.~Z.}\ \bibnamefont {Goodson}},
  \ and\ \bibinfo {author} {\bibfnamefont {D.~R.}\ \bibnamefont
  {Herschabach}},\ }\href@noop {} {\bibfield  {journal} {\bibinfo  {journal}
  {Theor. Chim. Acta}\ }\textbf {\bibinfo {volume} {81}},\ \bibinfo {pages} {1}
  (\bibinfo {year} {1991})}\BibitemShut {NoStop}%
\bibitem [{\citenamefont {Tsipis}\ \emph {et~al.}(1996)\citenamefont {Tsipis},
  \citenamefont {Popov}, \citenamefont {Herschbach},\ and\ \citenamefont
  {Avery}}]{avery_book}%
  \BibitemOpen
  \bibfield  {author} {\bibinfo {author} {\bibfnamefont {C.}~\bibnamefont
  {Tsipis}}, \bibinfo {author} {\bibfnamefont {V.}~\bibnamefont {Popov}},
  \bibinfo {author} {\bibfnamefont {D.}~\bibnamefont {Herschbach}}, \ and\
  \bibinfo {author} {\bibfnamefont {J.}~\bibnamefont {Avery}},\ }\href@noop {}
  {\emph {\bibinfo {title} {New Methods in Quantum Theory}}}\ (\bibinfo
  {publisher} {Kluwer Academic Publishing},\ \bibinfo {year} {1996})\ pp.\
  \bibinfo {pages} {33--54}\BibitemShut {NoStop}%
\bibitem [{\citenamefont {Nguyen}\ and\ \citenamefont
  {Nguyen-Dang}(2000)}]{dichotomy}%
  \BibitemOpen
  \bibfield  {author} {\bibinfo {author} {\bibfnamefont {N.~A.}\ \bibnamefont
  {Nguyen}}\ and\ \bibinfo {author} {\bibfnamefont {T.-T.}\ \bibnamefont
  {Nguyen-Dang}},\ }\href
  {http://scitation.aip.org/getpdf/servlet/GetPDFServlet?filetype=pdf&id=JCPSA6000112000003001229000001&idtype=cvips&doi=10.1063/1.480675&prog=normal}
  {\bibfield  {journal} {\bibinfo  {journal} {J. Chem. Phys.}\ }\textbf
  {\bibinfo {volume} {112}},\ \bibinfo {pages} {1229} (\bibinfo {year}
  {2000})}\BibitemShut {NoStop}%
\bibitem [{\citenamefont {Yasuike}\ and\ \citenamefont
  {Someda}(2004)}]{0953-4075-37-15-010}%
  \BibitemOpen
  \bibfield  {author} {\bibinfo {author} {\bibfnamefont {T.}~\bibnamefont
  {Yasuike}}\ and\ \bibinfo {author} {\bibfnamefont {K.}~\bibnamefont
  {Someda}},\ }\href {http://stacks.iop.org/0953-4075/37/i=15/a=010} {\bibfield
   {journal} {\bibinfo  {journal} {J. Phys. B}\ }\textbf {\bibinfo {volume}
  {37}},\ \bibinfo {pages} {3149} (\bibinfo {year} {2004})}\BibitemShut
  {NoStop}%
\bibitem [{\citenamefont {Rotenberg}\ \emph {et~al.}(2002)\citenamefont
  {Rotenberg}, \citenamefont {Taieb}, \citenamefont {Veniard},\ and\
  \citenamefont {Maquet}}]{0953-4075-35-17-103}%
  \BibitemOpen
  \bibfield  {author} {\bibinfo {author} {\bibfnamefont {B.}~\bibnamefont
  {Rotenberg}}, \bibinfo {author} {\bibfnamefont {R.}~\bibnamefont {Taieb}},
  \bibinfo {author} {\bibfnamefont {V.}~\bibnamefont {Veniard}}, \ and\
  \bibinfo {author} {\bibfnamefont {A.}~\bibnamefont {Maquet}},\ }\href
  {http://stacks.iop.org/0953-4075/35/i=17/a=103} {\bibfield  {journal}
  {\bibinfo  {journal} {J. Phys. B}\ }\textbf {\bibinfo {volume} {35}},\
  \bibinfo {pages} {L397} (\bibinfo {year} {2002})}\BibitemShut {NoStop}%
\end{thebibliography}%

\begin{center}
\newpage

\newpage

\begin{figure}
\centering
\includegraphics[width=18cm]{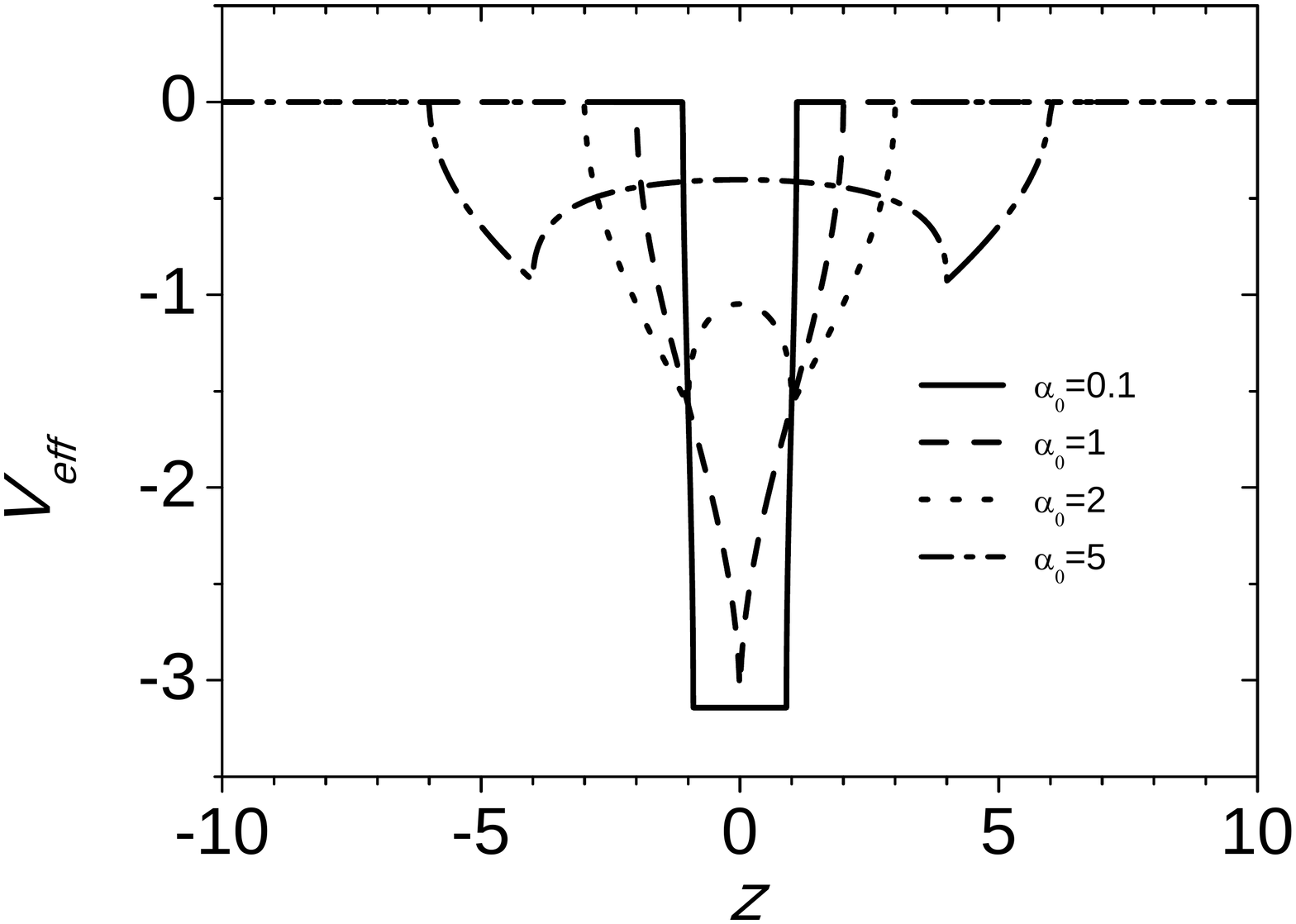}
\caption{Non-relativistic effective potential for a 1D particle in a box under different laser intensity, measured by $\alpha_0$. }
\label{nonrelativistic}
\end{figure}

\newpage

\begin{figure}
\centering
\includegraphics[width=8cm]{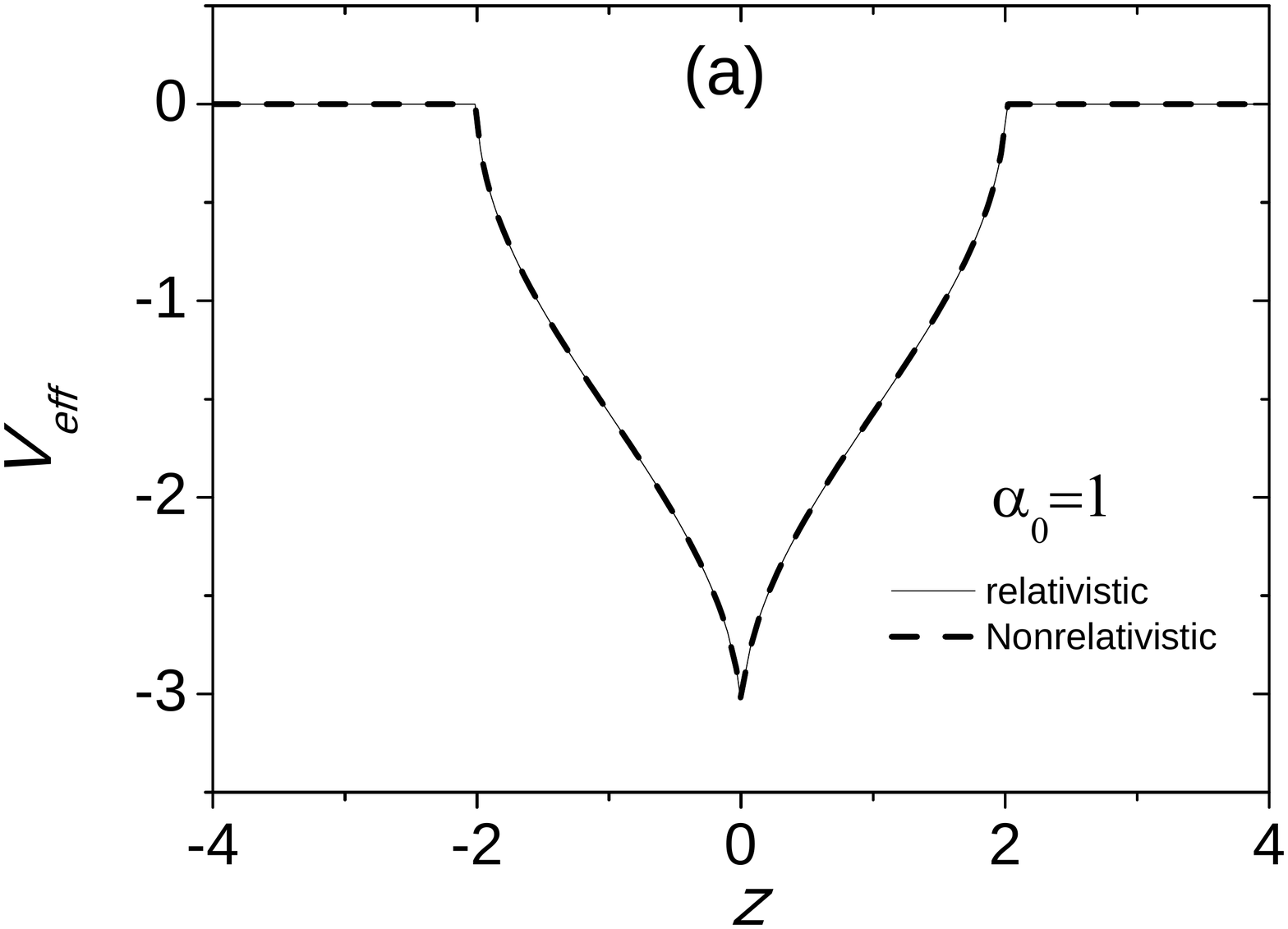}
\includegraphics[width=8cm]{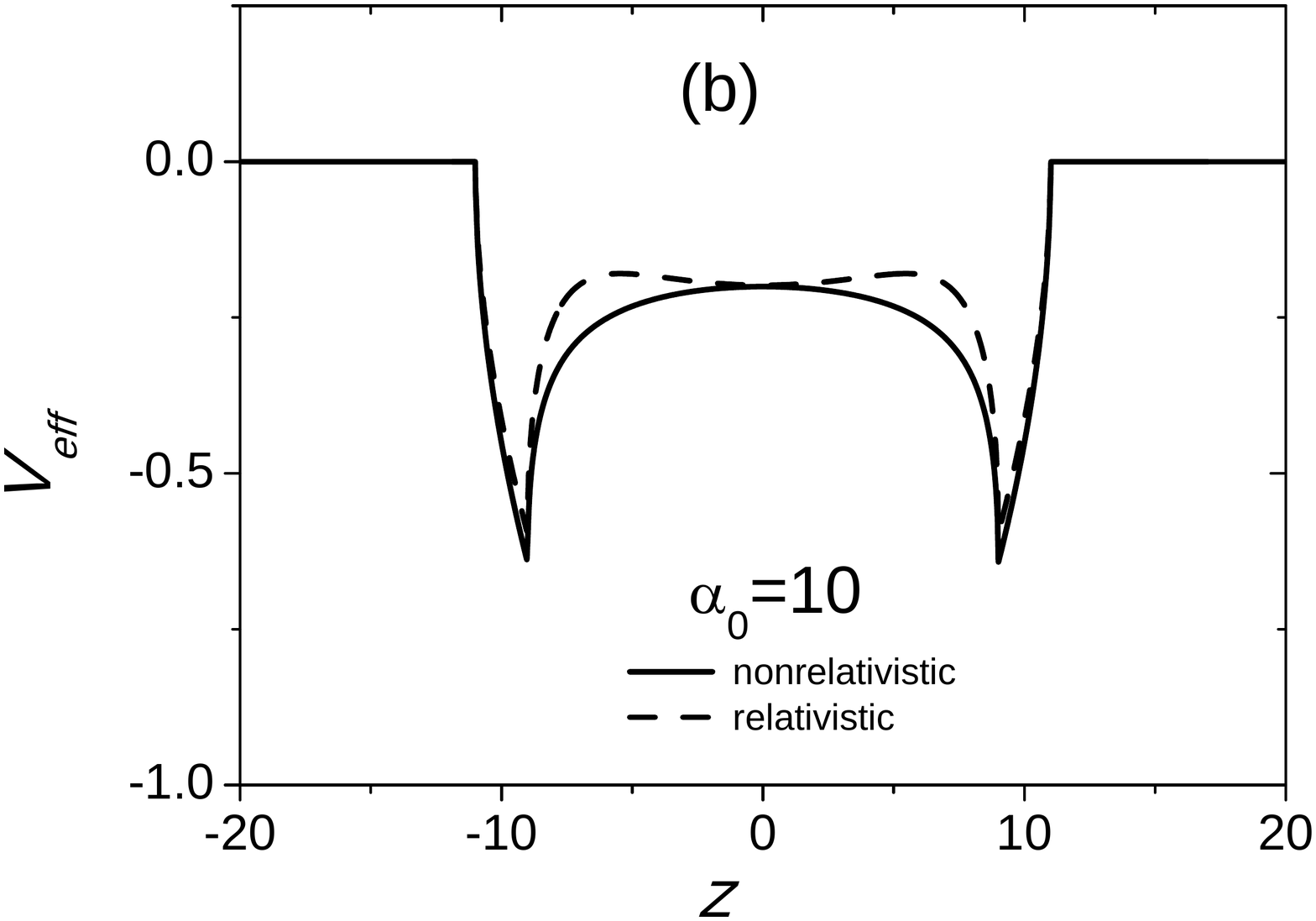}
\includegraphics[width=8cm]{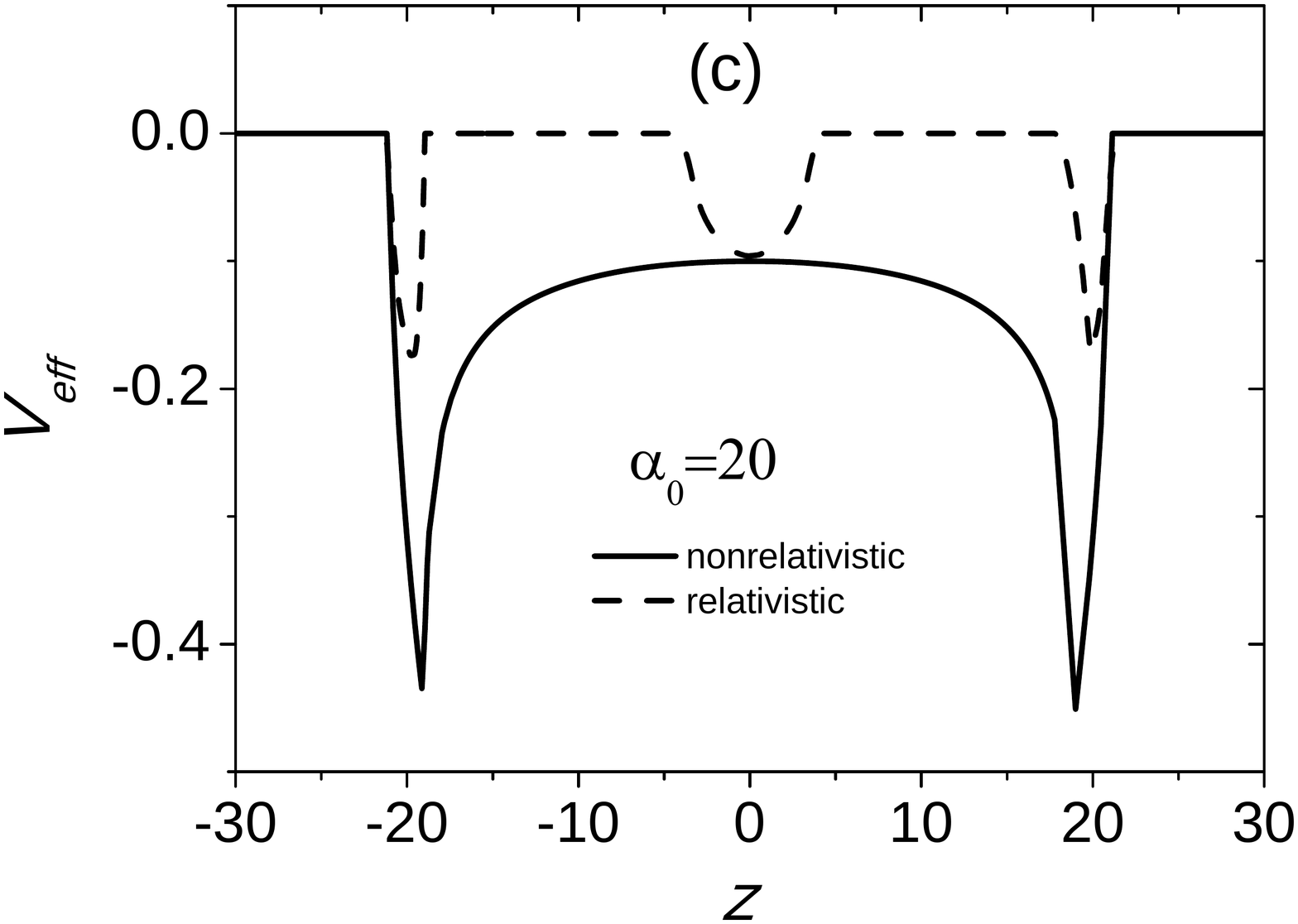}
\caption{Relativistic corrections to the effective potential for different laser fields.}
\label{relativistic}
\end{figure}

\newpage

\begin{figure}
\centering
\includegraphics[width=16cm]{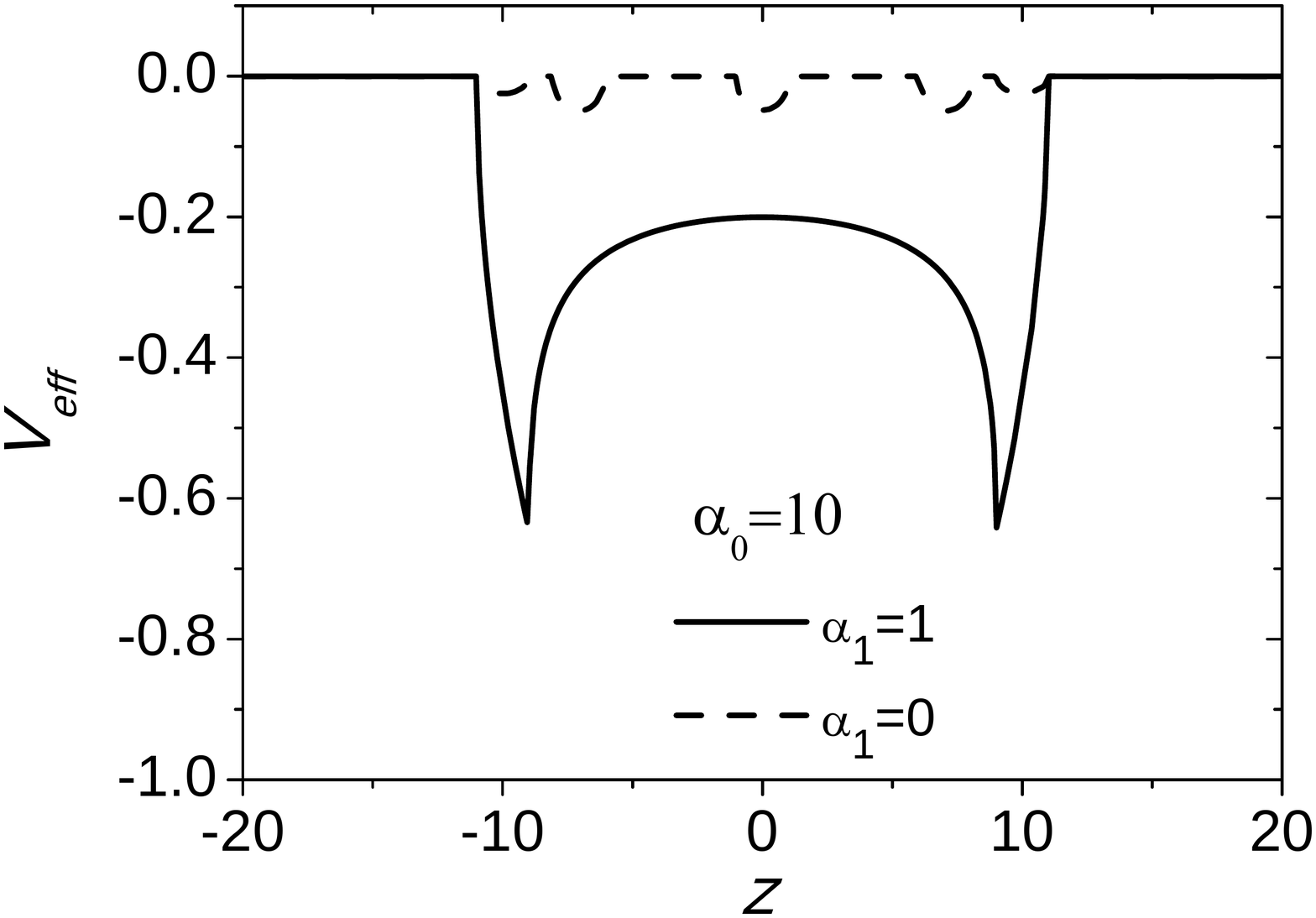}
\caption{Effective potential for two alternating electric fields superposed along $\hat e_x$ and $\hat e_z$, respectively, with different colors.}
\label{multicolor}
\end{figure}

\newpage

\begin{figure}
\includegraphics[scale=.8]{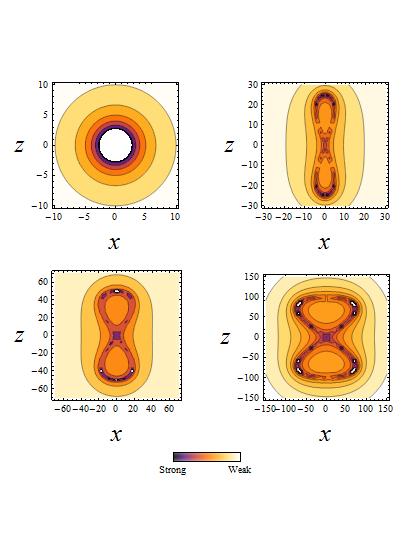}
\caption{Contour Plots of Dressed Potential for (clockwise) $\alpha_{0}$=0, 25, 100, 50.  Note both the shift in regime as $\alpha_{0}$ grows and the key below the plots for the interpretation of the intensity of the contours.}
\label{fig:contours}
\end{figure}

\newpage

\begin{figure}
\centering
\includegraphics[scale=.5]{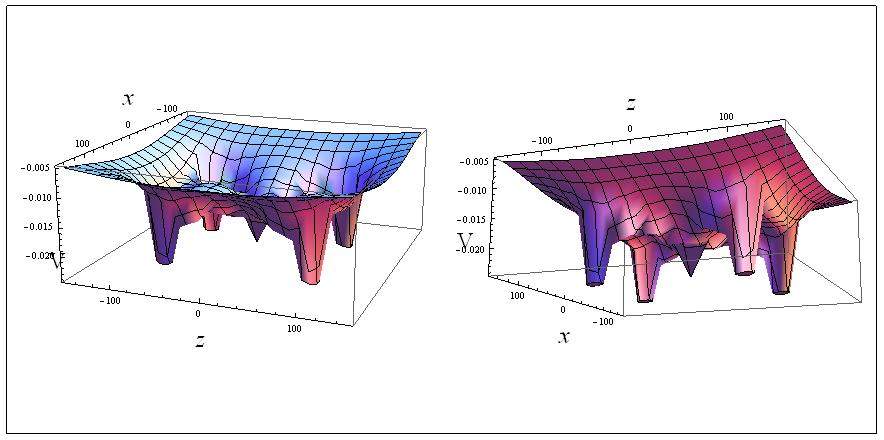}
\caption{Three Dimensional Plots of the Potential Energy, $V_{dres}^{HFFT}$, as a function of x and z coordinates for the case $\alpha_{0}$=100.  Left and Right of above are two different angles of same surface.}
\label{fig:surfaces}
\end{figure}

\newpage

\begin{figure}
\includegraphics[scale=.60]{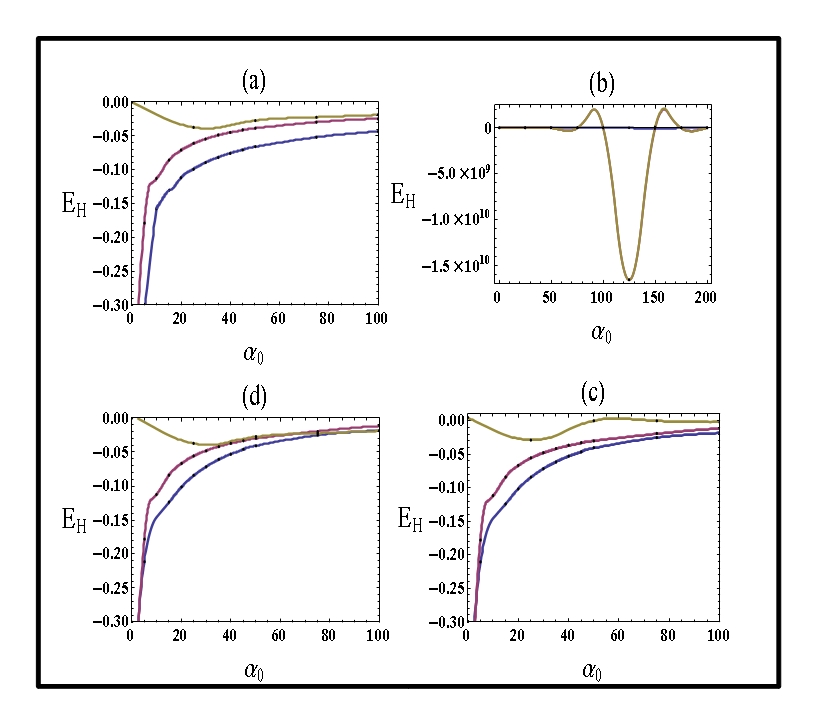}
\caption{Plots of both moecular energies and binding energy for (clockwise): $\mathcal{H}_{DA}$ with non-relativistic trajectory; $\mathcal{H}_{DA}$, relativistic trajectory; $\mathcal{H}_{P}$, relativistic trajectory; and $\mathcal{H}_{P}$ with non-relativistic trajectory.  All may be read as Yellow:Binding Energy; Blue:Hydrogen Energy; Purple:H$^{-}$ Energy }
\label{fig:Global}
\end{figure}

\newpage

\begin{figure}
\includegraphics[scale=.45]{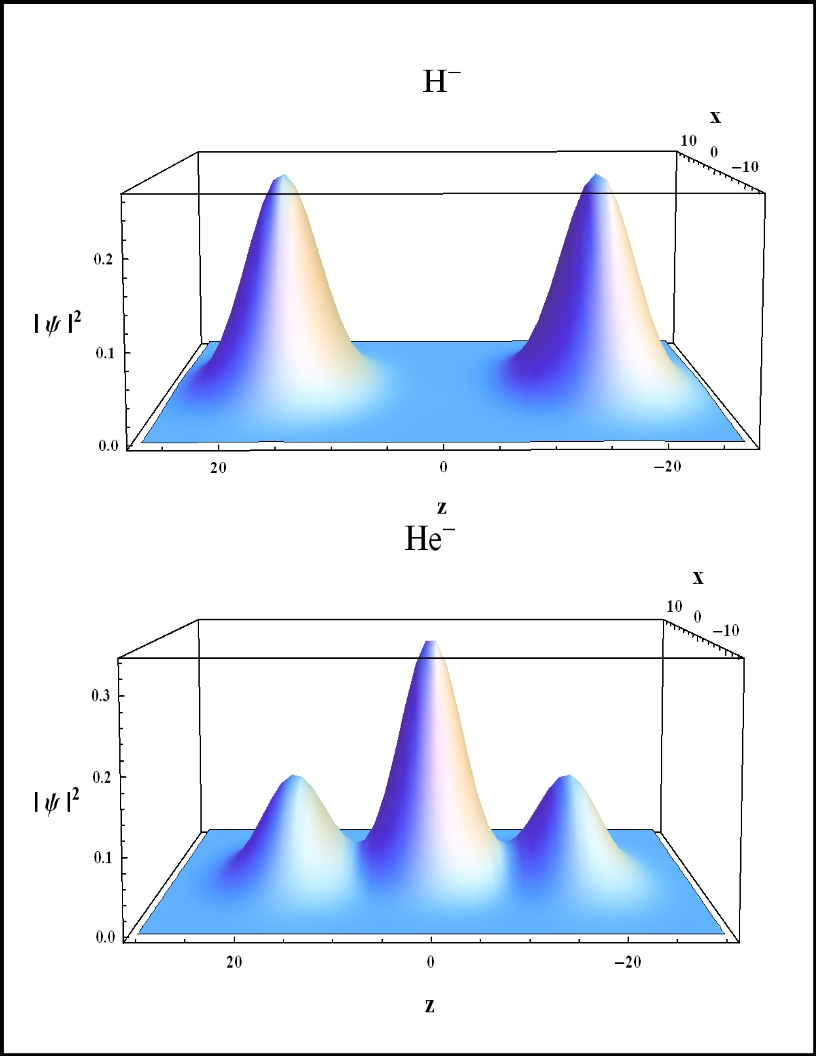}
\caption{Top: Probability distribution for H$^{-}$, a two electron system.  Bottom: Probability distribution for He$^{-}$, a three electron system}
\label{fig:lastpix}
\end{figure}

\newpage

\begin{figure}
\includegraphics[scale=.65]{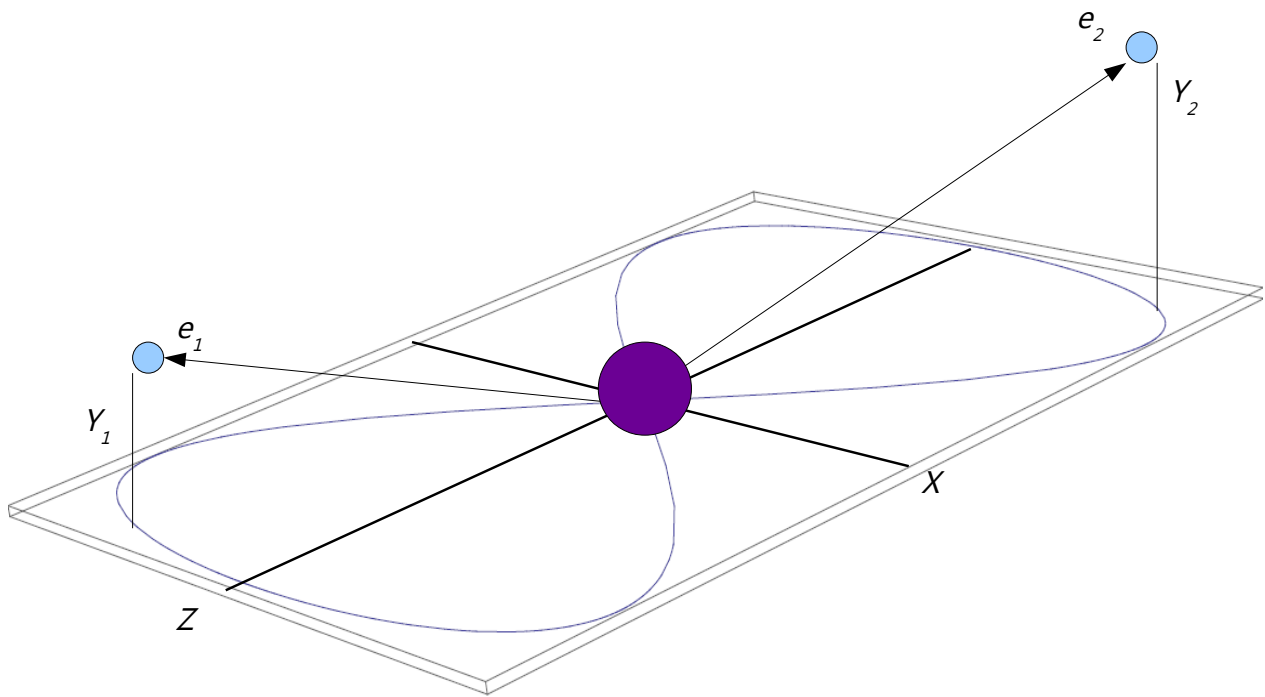}
\caption{The above displays the relationship between the system's geometry with respect to the electrons and the coordinates use in Eq. (\ref{eq:new})}
\label{fig:coords}
\end{figure}

\newpage

\begin{figure}
\includegraphics[scale=.3]{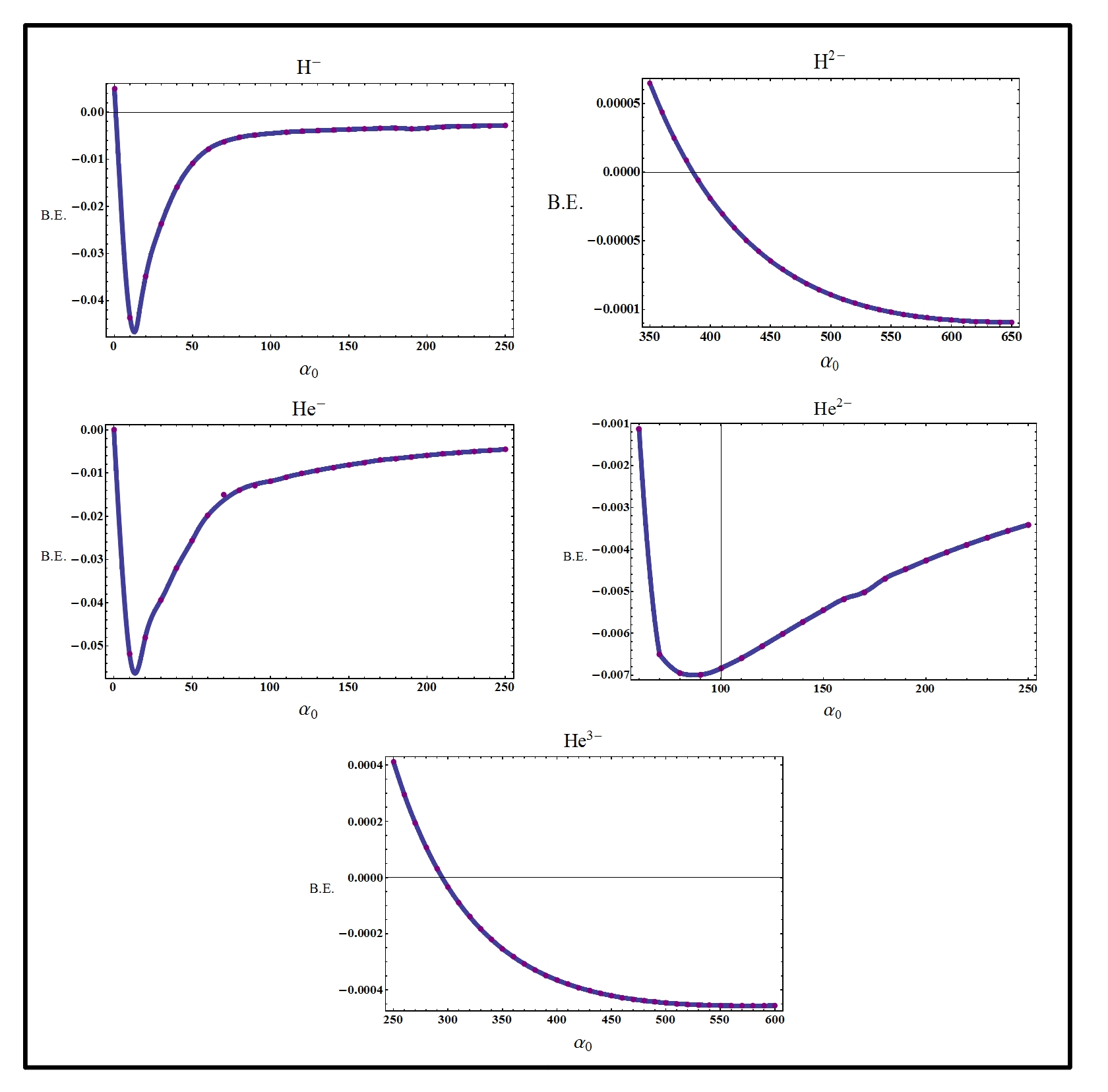}
\caption{ Above are plots of the binding energies of,  from left to right and top to bottom: H$^{-}$, H$^{-2}$, He$^{-}$, He$^{-2}$ and He$^{-3}$.}
\label{fig:BEs}
\end{figure}

\newpage

\begin{figure}
\includegraphics[scale=.4]{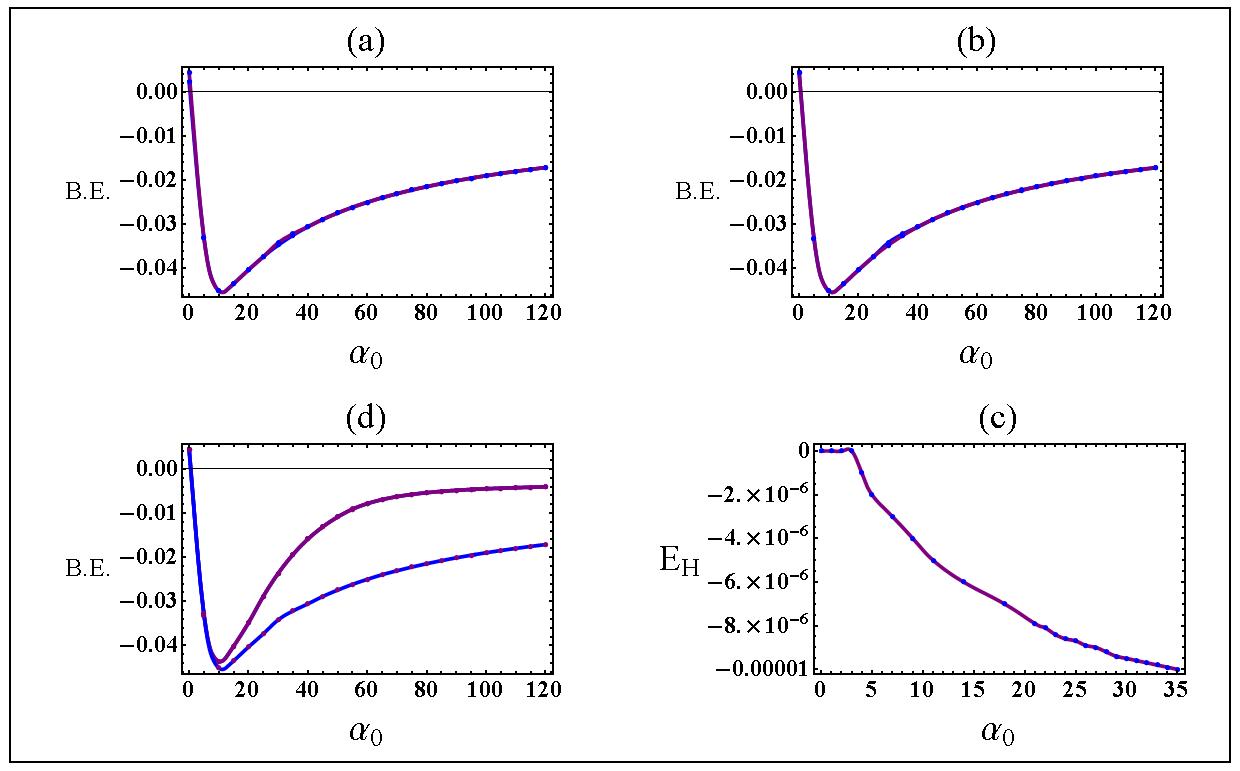}
\caption{Plots of binding energy comparisons for (clockwise):comparison plot of $\mathcal{H}_{DA}$ versus $\mathcal{H}_{P}$, non-relativistic trajectory; $\mathcal{H}_{P}$ with relativistic trajectory m$_{0}$ and m$_{r}$, differences value of Hydrogen energy between use of m$_{0}$ and m$_{r}$; and a comparison of relativistic trajectory (both m$_{0}$ and m$_{r}$) with non-relativistic trajectory both using $\mathcal{H}_{P}$.}
\label{fig:Local}
\end{figure}

\newpage

\begin{figure}
\includegraphics[scale=.33]{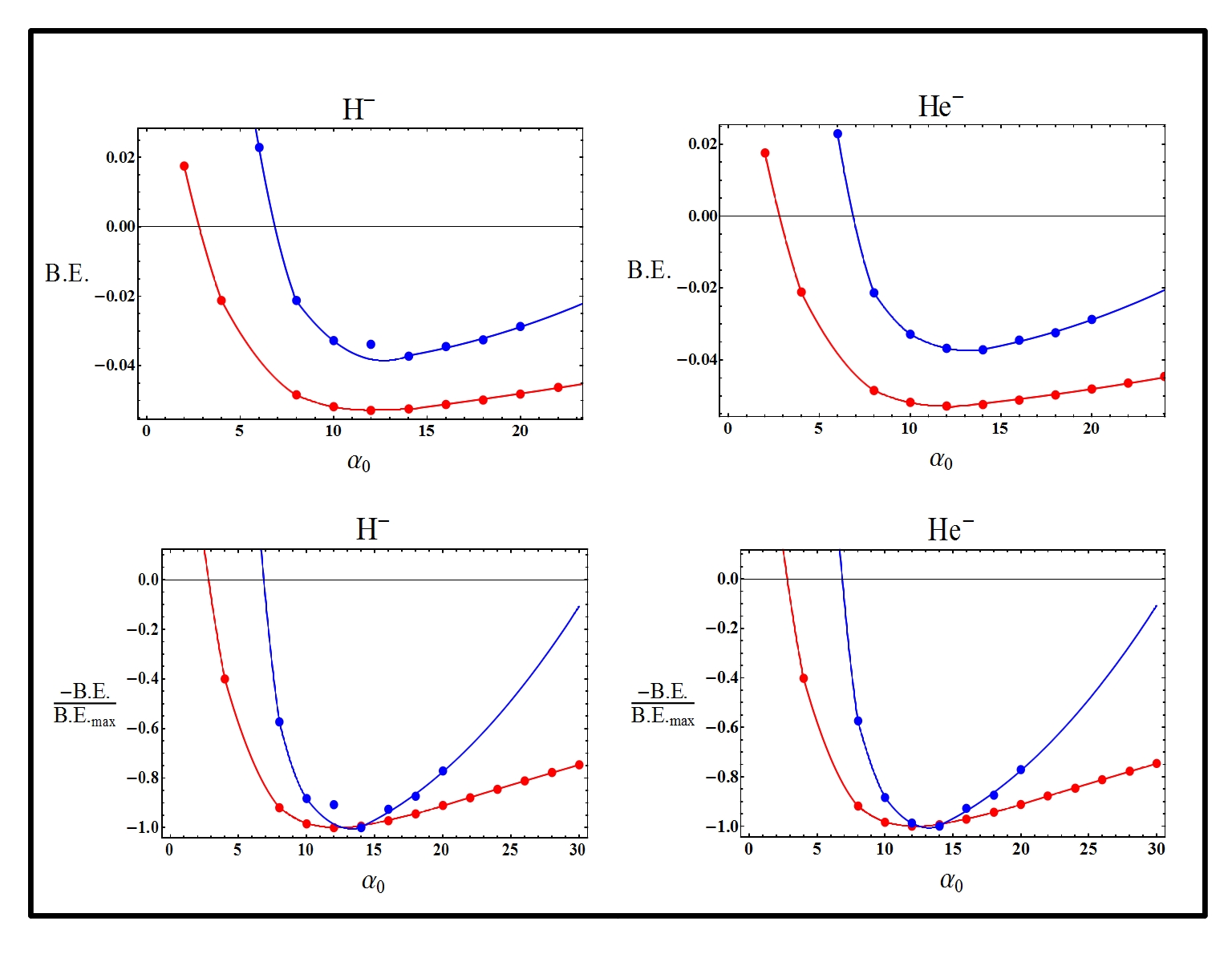}
\caption{Plots of binding energy for the H$^{-}$ (left) and He$^{-}$ (right) systems.  Top, Non-Normalized plots of the calculation data showing agreement between the methods.  Bottom, -B.E./B.E.$_{max}$ to emphasis the qualitative simularity between the methods as they share minima for the B.E. curves.}
\label{fig:BEs3D}
\end{figure}

\newpage

\begin{figure}
\includegraphics[scale=.25]{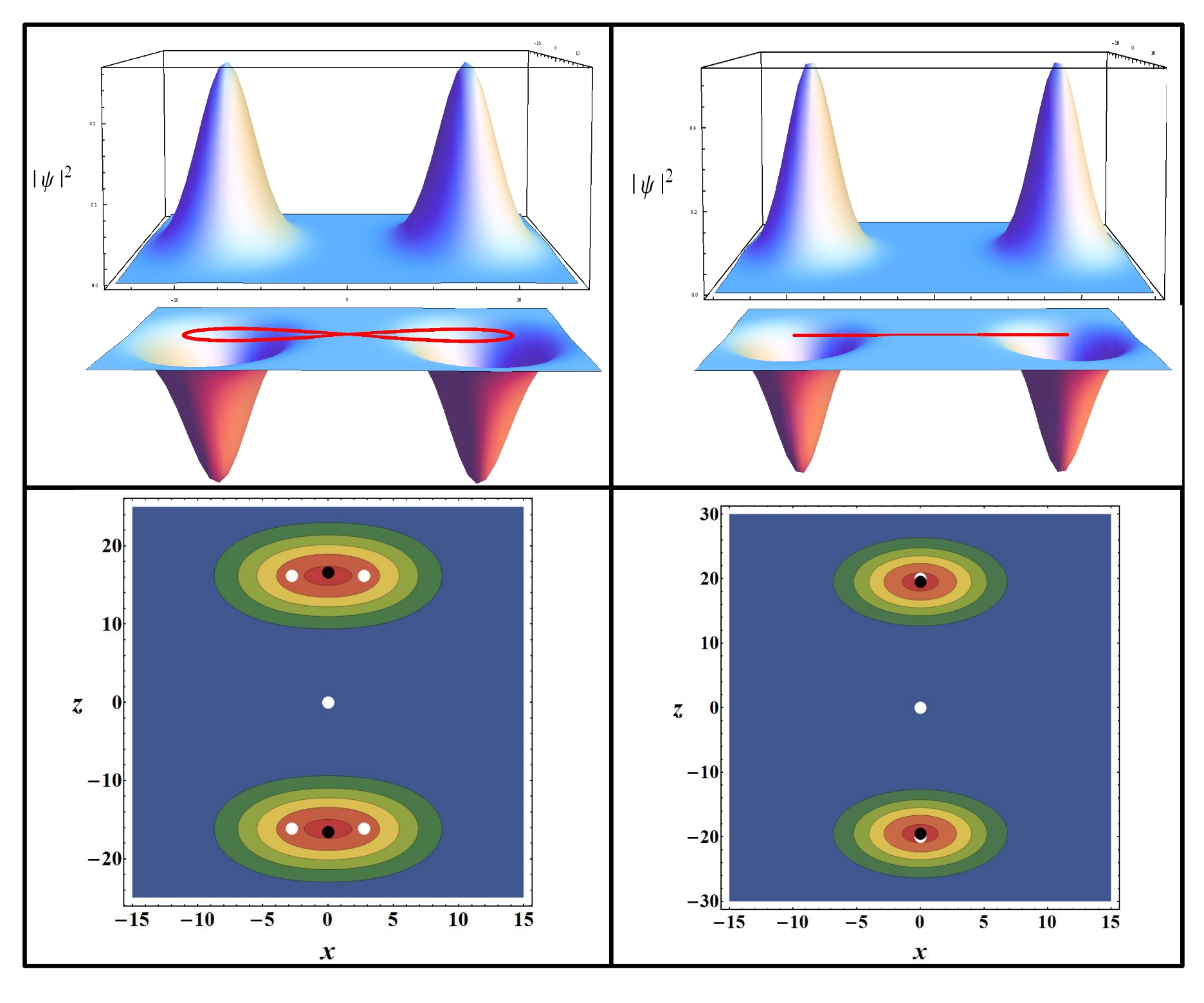}
\caption{Top Row: Plots of the Probabiltiy Distribution for the corrected (left) and non-corrected (right) H$^{-}$ system, directly below is a superimposition of the trajectory upon the the probabilty density function plot to emphasis their relation.  Bottom Row:  Contour Plots of the H$^{-}$ system, both corrected (left) and non-corrected (right), note the different scales on the vertical (z) axis and the more diffuse behavior of the corrected system.}
\label{fig:probconts}
\end{figure}

\newpage

\begin{figure}
\includegraphics[scale=.5]{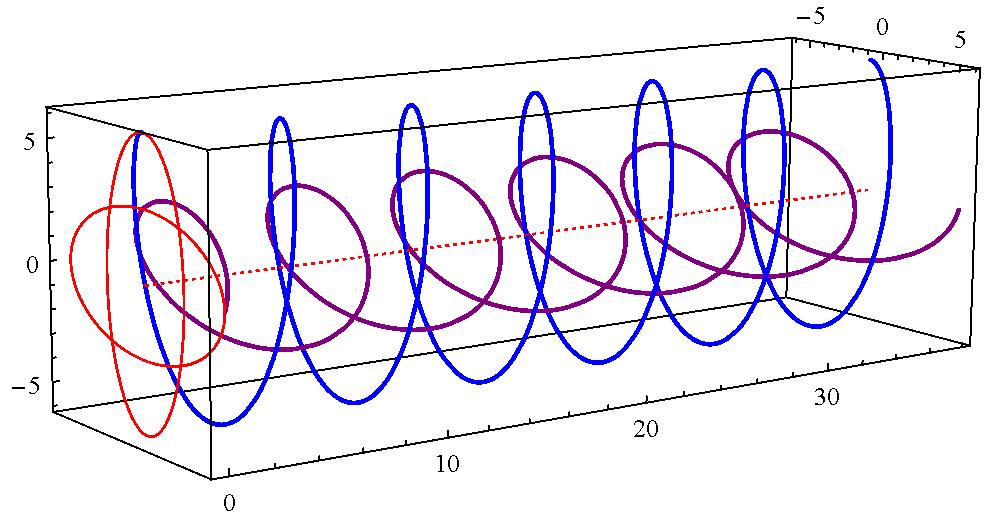}
\caption{The electronic elliptical contribution can be seen as the function whose major axis lies in the z-direction (vertical) and the magnetic contribution has an orthogonal orientation.}
\label{fig:SeperateEll}
\end{figure}

\newpage

\begin{figure}
\includegraphics[scale=.45]{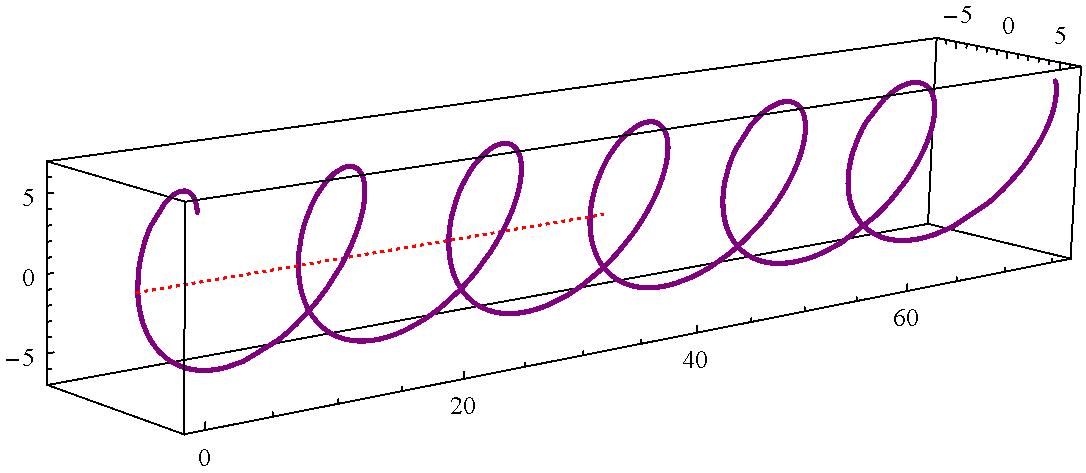}
\caption{Within the above plot, the total trajectory can be seem, the amplitude of the major and minor axises are mediated in value between those from the electronic and magnetic components and the orientation is set off by an angle whose value respects the same coefficients as the relative amplitudes.}
\label{fig:combinedEll}
\end{figure}

\newpage

\begin{table}[h]
  \begin{tabular}{| l | l | l | l | }
    \hline
    Species & Inner Oribital& Outer Orbital&Total Electron\\
    & Population&Population (each)&Count (N) \\ \hline
    H & 0.01645 & 0.24576 &1 \\ \hline
    H$^{-}$ & 0.000524 & 0.49987 & 2 \\ \hline
    He & 0.000432 & 0.499892 &2 \\ \hline
    He$^{-}$&1.00001&0.499999&3 \\
    \hline
    \end{tabular}
  \caption{Results of Mulliken Population Analysis, note that there are four outer orbitals the table contains one of the four values.}
  \label{my_table}
\end{table}

\end{center}

\end{document}